\documentclass[aps,pra,reprint,amssymb]{revtex4-1}

\usepackage{amsmath}
\usepackage{bm}
\usepackage{graphicx}
\usepackage{braket}
\usepackage[colorlinks=true, linkcolor=blue]{hyperref} 
\usepackage{mathtools}

\newcommand{\R}{\mathbb{R}}

\renewcommand{\i}{\mathrm{i}}
\newcommand{\hamilton}{\hat{\mathcal{H}}}

\newcommand{\vectx}{\vect{x}}
\newcommand{\vectk}{\vect{k}}

\newcommand{\ie}{i.e.}
\newcommand{\eg}{e.g.}
\renewcommand{\Re}{\operatorname{Re}}
\newcommand{\rt}{\tilde{r}}
\newcommand{\ct}{\tilde{c}}

\DeclareMathOperator{\erf}{erf}
\DeclareMathOperator{\erfc}{erfc}

\DeclareMathOperator{\sgn}{sgn}

\DeclarePairedDelimiter{\NormalO}{:}{:} 

\newcommand*{\Diff}[2]{\mathop{}\!\mathrm{d}^{#1}{#2}\,}
\newcommand*{\diff}[1]{\mathop{}\!\mathrm{d}{#1}\,}
\newcommand{\invLaplace}[3]{\mathcal{L}_{#1}^{-1}\Big[#2\Big](#3)}

\newcommand{\eh}[1]{\mathrm{e}^{#1}}
\newcommand{\vect}[1]{\mathbf{#1}}

\newcommand{\secondderivative}[1]{\frac{\partial^2}{\partial {#1}^2}}
\newcommand{\ursell}[1]{\hat{U}^{(#1)}}
\newcommand{\K}[1]{K^{(#1)}}
\newcommand{\Kop}[1]{\hat{K}^{(#1)}}
\newcommand{\dK}[1]{\Delta K^{(#1)}}

\begin{document}

\title{Nonlocal pair correlations in Lieb-Liniger gases: \\a unified non-perturbative approach from weak degeneracy to high temperatures}

\author{Benjamin Geiger}
\email[Corresponding author: ]{benjamin.geiger@ur.de}
\affiliation{Institut f\"{u}r Theoretische Physik, Universit\"{a}t Regensburg, D-93040 Regensburg, Germany}

\author{Quirin Hummel}
\affiliation{Institut f\"{u}r Theoretische Physik, Universit\"{a}t Regensburg, D-93040 Regensburg, Germany}
\author{Juan Diego Urbina}
\affiliation{Institut f\"{u}r Theoretische Physik, Universit\"{a}t Regensburg, D-93040 Regensburg, Germany}
\author{Klaus Richter}
\affiliation{Institut f\"{u}r Theoretische Physik, Universit\"{a}t Regensburg, D-93040 Regensburg, Germany}
\date{June 15, 2018}

\begin{abstract}
We present analytical results for the nonlocal pair correlations in one-dimensional bosonic systems with repulsive contact interactions that are uniformly valid from the classical regime of high temperatures down to weak quantum degeneracy entering the regime of ultralow temperatures. By using the information contained in the short-time approximations of the full many-body propagator, we derive results that are nonperturbative in the interaction parameter while covering a wide range of temperatures and densities. For the case of three particles we give a simple formula for arbitrary couplings that is exact in the dilute limit while remaining valid up to the regime where the thermal de Broglie wavelength $\lambda_T$ is of the order of the characteristic length $L$ of the system. We then show how to use this result to find analytical expressions for the nonlocal correlations for arbitrary but fixed particle numbers $N$ including finite-size corrections. Neglecting the latter in the thermodynamic limit provides an expansion in the quantum degeneracy parameter $N\lambda_T/L$. We compare our analytical results with numerical Bethe ansatz calculations, finding excellent agreement.
\end{abstract}

\maketitle

\section{INTRODUCTION}
The study of spatial correlations provides an intuitive and experimentally accessible window to the physical properties of interacting many-body quantum systems. The special role of low-order spatial correlation functions arises from the definitional property of multiparticle systems as having a large number of degrees of freedom. Up to the case of two or three degrees of freedom, the spatial structure of the wave function can be directly visualized and efficiently computed. When the number of degrees of freedom increases, the full description of quantum-mechanical states not only becomes highly unintuitive, but pretty soon explicit computations become a hopeless task. This is one of the reasons for the relevance of field-theoretical descriptions in terms of field operators that live in real space and provide more intuitive characterizations in terms of collective degrees of freedom such as particle density and correlation functions \cite{Negele1998}. These theoretical descriptions have been used to successfully describe quantities accessible to measurements in noninteracting ultracold atom systems \cite{Yasuda1996,Ottl2005,Schellekens2005,Jeltes2007,Manz2010,Manning2010,Hodgman2011,Rugway2013,Dall2013,Schweigler2017}.

For interacting systems state-of-the-art experiments \cite{Kinoshita2005,Vogler2013} have addressed so far mainly the local limit $g_2(\vect{r}\to 0)$ of the (normalized) pair-correlation function 
\begin{equation}
g_2(\vect{r})=\frac{\braket{\hat{\Psi}^\dagger(0)\hat{\Psi}^\dagger(\vect{r})\hat{\Psi}(\vect{r})\hat{\Psi}(0)}}{\braket{\hat{\Psi}^\dagger(0)\hat{\Psi}(0)}\braket{\hat{\Psi}^\dagger(\vect{r})\hat{\Psi}(\vect{r})}},
\label{eqn:ggeneral}
\end{equation}
here expressed in terms of the bosonic field operators $\hat{\Psi}$ and $\hat{\Psi}^\dagger$ [see also \cite{Haller2011} for recent results on $g_3(0)$], while specific proposals for the measurement of truly nonlocal correlations with $\vect{r}\neq0$ are now available \cite{Elliott2016,Streif2016}.

Within the program of characterizing the spatial structure of many-body states, one-dimensional (1D) systems play a special role.
One reason for this is the possibility of experimental realization \cite{Kinoshita2004,Paredes2004}, where now controlled access to the collective behavior of a few dozens of constituents is possible \cite{Bernien2017}. Moreover, for this kind of systems, and depending on the type of interaction and other properties, the corresponding mathematical description may fall into the category of quantum integrable models and thus admits an explicit (but formal) solution in terms of a set of algebraic equations.
A paradigmatic example of quantum integrability is the Lieb-Liniger model \cite{lieb1963}, a many-body Hamiltonian describing a set of $N$ bosonic particles interacting through repulsive short-range forces, and confined to a region of finite length $L$.
One of the remarkable consequences of quantum integrability is that the many-body  eigenstates and eigenenergies of these systems are characterized by a complete set of quantum numbers labeling the rapidities of the states \cite{lieb1963,gaudin2014}.
The latter, although playing the role of quasimomenta, are, however, genuine many-body objects that do not have a direct interpretation in terms of quasiparticle excitations unless the particle number becomes infinite \cite{Lieb1963b}.

Although the theory of quantum integrable systems provides, in principle, results for any kind of spatial correlations to any order \cite{Zill2016}, it has two obvious drawbacks.
First, the solutions of the equations relating the quantum numbers to the actual quantized quasimomenta must be found numerically, even for the case of two particles, and becomes more and more a black-box routine when the regime of a few to dozens of particles is reached.
Second, in finite systems where finite temperatures enter into consideration, the usefulness of precise quantized many-body eigenstates is even more questionable, as one expects the many-body spectra to get exponentially dense \cite{Bethe1936}.
These problems stem from the discrete character of the Bethe ansatz equations.
Usually, one considers the thermodynamic limit to overcome them in what is known as thermodynamic Bethe ansatz \cite{yang1969} or by exploiting the asymptotic equivalence to grand-canonical descriptions.
However, besides the obvious limitation to very large particle numbers, related approaches to address nonlocal multiparticle correlations also suffer from restrictions to the extreme regimes of weak or strong coupling~\cite{Nandani2016,Deuar2009} and small interparticle separations \cite{Nandani2016}.

In this paper, a different approach to spatial correlations in interacting quantum systems in thermal equilibrium will be presented that is especially useful in the few-particle regime. The underlying concept is based on the fact that for finite temperatures and for the whole range of interaction strengths, the discreteness of the multiparticle spectrum due to spatial confinement cannot be resolved except in the quantum degenerate regime. Therefore we assume that only short-time information, \ie, approximating the many-body dynamics by its bulk contribution with smoothed spectrum, should provide the major physical input. Once this point of view is adopted, the difficulty consists in expressing quantities of physical interest in terms of short-time processes. This will be done using the standard tool of cluster expansions \cite{Kahn1938,Gruter1995,Gruter1995b,Gruter1997}.

We note that, within our approach, interaction effects are treated fully nonperturbatively in the short-time approximation, and therefore our results will cover the entire range of interaction strengths within the regime where the discreteness of the many-body spectrum can be neglected. This is to be contrasted to perturbative or strong-coupling expansions, valid only near the limits of non- or strongly interacting systems \cite{Deuar2009}.

Our work is inspired by state-of-the-art experimental measurements of nonlocal pair correlations in ultracold $\mathrm{He}_4$ atomic clouds in quasi-1D geometries, as discussed in \cite{Dall2013}.
In this pioneering experiment, high-order nonlocal correlators are measured,
with the two-body correlation showing a Gaussian profile as a function of the separation, a clear indication of temperatures well above deep quantum degeneracy and negligible interactions. The validity of the measurement protocol in this nearly ideal Bose gas was additionally confirmed by the compatibility of measured high-order correlations with Wick's theorem, bringing nonlocal multiparticle correlations in \textit{interacting} quantum gases closer to experimental reach.
The approach presented here works well precisely in the regime of weak degeneracy, where (thermal) boson bunching is still strongly pronounced but already starts to decay into long-range coherence present in the BEC regime \cite{Ottl2005,Hodgman2011}.
By providing accurate unified analytical formulas in the whole range from weak to strong interactions we capture all their nontrivial effects on the bunching behavior in a single strike.

The paper is organized as follows: In Sec.~\ref{sec:ursell_and_clusters} we present the quantum cluster expansion using Ursell operators and the resulting expression for the nonlocal pair-correlation function. We also introduce the general properties of the short-time approximation. In Sec.~\ref{sec:LL} we apply the methods of the previous section to the Lieb-Liniger gas representing quasi-1D cold atoms in ring traps.
The power of the short-time approximation when combined with cluster expansions is tested against full-fledged numerical calculations based on the Bethe ansatz equations that solve the quantum integrable model. We provide closed analytical results for the nonlocal pair-correlation function for the whole temperature regime down to weak quantum degeneracy and valid for the full regime of interactions, including the extreme limit of fermionization.
\section{URSELL OPERATORS AND THE CLUSTER EXPANSION}
\label{sec:ursell_and_clusters}
\subsection{Ursell operators}
The method that we use for our calculations is quite general, hence we do not have to restrict ourselves to 1D systems or specific interaction potentials at this stage, as long as the latter are sufficiently short ranged. We assume that the particle number $N$ is fixed and that the system is in thermal equilibrium with its environment. The thermodynamic properties of the system are then fully described by the heat kernel $\braket{\vectx'|\eh{-\beta\hamilton}|\vectx}$, where $\hamilton$ is the $N$-particle Hamiltonian of the system, $\beta=(k_BT)^{-1}$ is the inverse temperature, and $\ket{\vectx}=\ket{\vectx_1}\otimes\dots\otimes\ket{\vectx_N}=\ket{\vectx_1,\dots,\vectx_N}$ is a product of $N$ position eigenstates. We can represent the heat kernel by the many-body propagator
\begin{equation}
\K{N}(\vectx',\vectx;t)=
\braket{\vectx'|\eh{-\frac{\i t}{\hbar}\hamilton}|\vectx},
\end{equation}
evaluated at imaginary time
\begin{equation}
	t=-\i\hbar\beta.
\end{equation}
For indistinguishable particles we have to use the symmetry projected equivalent,
\begin{equation}
	\K{N}_\pm(\vectx',\vectx;t)=\frac{1}{N!}\sum_{P\in S_N}(\pm 1)^P\K{N}(P\vectx',\vectx;t),
	\label{eqn:symmetric_propagator}
\end{equation}
where the sum runs over the symmetric group $S_N$ operating on the particle indices, $+$ and $-$ stand for bosons and fermions, and $(-1)^P$ is the sign of the permutation $P$.

To pave the way to approximate the propagator for distinguishable particles we decompose the imaginary-time evolution operator into Ursell operators \cite{Gruter1995} in the following manner. Let $\hamilton(i_1,\dots,i_n)$ be the part of the Hamiltonian that acts only on $n\leq N$ particles $i_1,\dots,i_n$ and $\Kop{n}(i_1,\dots,i_n)=\eh{-\frac{\i t}{\hbar}\hamilton(i_1,\dots,i_n)}$. The first three Ursell operators $\ursell{n}$ are then implicitly defined as 
\begin{align}
	\Kop{1}(1)&= \ursell{1}(1),\nonumber\\
	\Kop{2}(1,2)&=\ursell{1}(1)\ursell{1}(2)+\ursell{2}(1,2),\nonumber
	\\
	\Kop{3}(1,2,3)&=
	\ursell{1}(1)\ursell{1}(2)\ursell{1}(3)\nonumber
	\\&+
	\ursell{1}(1)\ursell{2}(2,3)+\ursell{1}(2)\ursell{2} (1,3)\nonumber
	\\&+
	\ursell{1}(3)\ursell{2}(1,2)
	+\ursell{3}(1,2,3)
.
\label{eqn:ursell}
\end{align}
All higher Ursell operators are defined in the same way by decomposing $\hat{K}^{(n)}$ into all possible particle partitions. Due to the short-range character of the interaction, particles that are separated far from each other will be essentially independent. This means that the matrix elements 
\begin{equation}
	\dK{n}(\vectx',\vectx;t)\equiv\braket{\vectx'|\ursell{n}|\vectx}
	\label{eqn:interaction_contribution}
\end{equation}
in coordinate space vanish if the distance of any two particles in $\vectx$ and $\vectx'$ is large. The propagator $\K{n}(\vectx',\vectx;t)$ can then be written in terms of the matrix elements $\dK{j}(\vectx',\vectx;t)$ with $j\leq n$. We will further refer to these matrix elements as interaction contributions of order $j$ and identify $\K{1}(\vectx',\vectx;t)=\dK{1}(\vectx',\vectx;t)$ for $j=1$. We can now write the propagator for $N$ distinguishable particles as a sum of interaction contributions
\begin{equation}
	\K{N}(\vectx',\vectx;t)=\sum_{\mathcal{J}\vdash\{1,\dots,N\}}\prod_{I\in\mathcal{J}}\dK{|I|}(\vectx'_I,\vectx_I;t),
	\label{eqn:cluster_expansion}
\end{equation}
where the sum in this cluster expansion runs over all possible partitions $\mathcal{J}$ of the $N$ particles and $\vectx_I$ is the shorthand notation for all particle coordinates that are part of the same interaction contribution. This decomposition is particularly useful when higher-order interaction contributions are subdominant, \ie, the dominant parts of the propagator factorize into clusters of smaller particle numbers. We stress that neglecting, \eg, interaction contributions of order $n\geq 3$ is conceptually different from a perturbation expansion, as two-body interactions are fully accounted for by the interaction contributions of order $n=2$, which are nonperturbative in the interaction strength. While respecting the finiteness of the system, such a truncation includes the virial expansion to second order in the thermodynamic limit.

In the case of indistinguishable particles there is an additional factorization mechanism corresponding to the decomposition of permutations into cycles \cite{hummel2013}. This naturally leads to a grouping of particles in clusters that are either part of the same interaction contribution or connected by permutation cycles. This becomes important when calculating traces of the propagator, as each cluster of particles can then be treated independently from the rest of the particles while its internal dynamics is tied in a nonseparable way. As an illustrative example, consider a partition of $N\geq 3$ particles into one interaction contribution of order 2 [\eg, particles one and two connected by $\ursell{2}(1,2)$] and $N-2$ interaction contributions of order 1, together with the permutation $P=(1~3)$. This is one of many combinations that appear if we symmetrize Eq.~(\ref{eqn:cluster_expansion}) according to Eq.~(\ref{eqn:symmetric_propagator}). It factorizes into $N-3$ single-particle propagators and the term
\begin{equation}
	\dK{2}\boldsymbol{(}(\vectx_3',\vectx_2'),(\vectx_1,\vectx_2);t\boldsymbol{)}\K{1}(\vectx_1',\vectx_3;t).
	\label{eqn:3-cluster}
\end{equation}
An additional factor $1/N!$ in Eq.~(\ref{eqn:symmetric_propagator}) accounts for the correct normalization. So, in this example we have a total of $N-2$ clusters---one cluster comprising three particles and $N-3$ (trivial) single-particle clusters. Even though the factors in Eq.~(\ref{eqn:3-cluster}) are, as is, independent functions, they cannot be treated independently if we trace, \eg, the particle with index 3, showing that the relevant criterion of factorization into independent clusters is the particle index rather than the coordinates themselves.
\subsection{Diagrams}
In order to calculate thermodynamic quantities or reduced density matrices one has to (partially) trace the $N$-particle propagator. Already for moderate particle numbers this leads to a plethora of identical contributions in Eq.~(\ref{eqn:cluster_expansion}) due to particle relabeling. This suggests a diagrammatic treatment of the (symmetry-projected) cluster expansion (\ref{eqn:cluster_expansion}). Each interaction contribution of order $n$ is thus represented as a diagram connecting $n$ initial and $n$ final coordinates.
\begin{figure}
	\includegraphics[width=\linewidth]{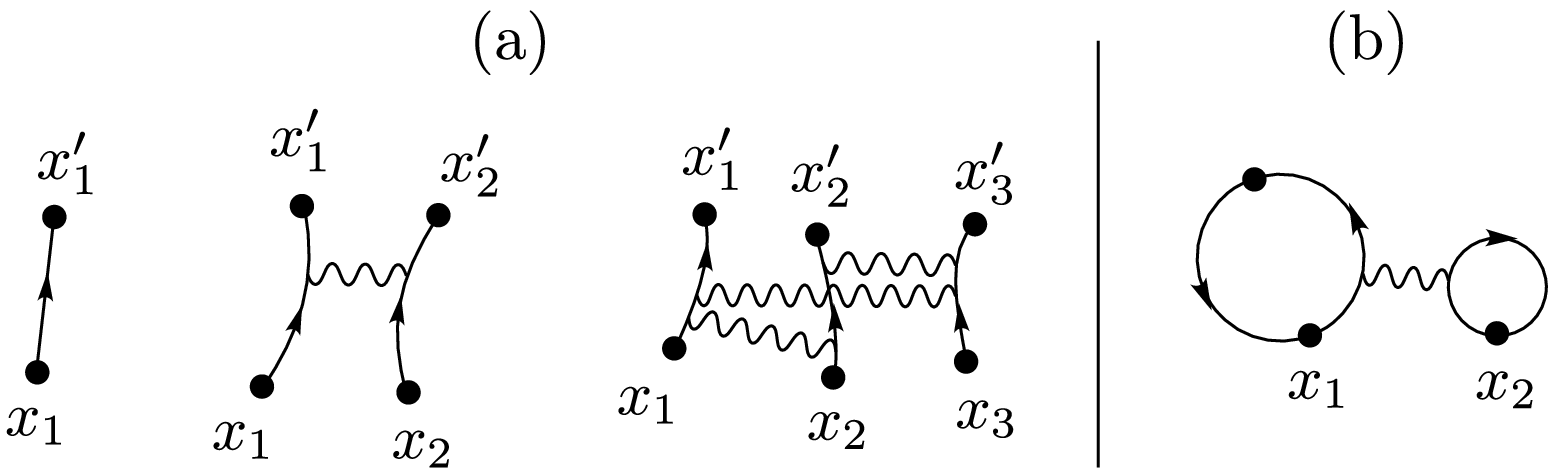}
	\caption{(a) Diagrams representing $\dK{n}(\vectx',\vectx;t)$, Eq.~(\ref{eqn:interaction_contribution}), for $n=1,2,3$. (b) Diagram representing the particular cluster Eq.~(\ref{eqn:3-cluster}) for $\vectx=\vectx'$.}
	\label{fig:diagrams}
\end{figure}
The diagrams for the first three orders are displayed in Fig.\@ \ref{fig:diagrams}(a), where the particle coordinates are marked by labeled dots. Diagrams that appear in (partial) traces are constructed from those building blocks. A full diagram represents a factorization into clusters according to Eq.~(\ref{eqn:cluster_expansion}) or its symmetry-projected equivalent and comprises several irreducible diagrams that represent single clusters. By convention each unlabeled bullet in a diagram stands for a coordinate that has been traced out. Such points have to be connected to two other points in the diagram. Loose ends are possible in general (for example in off-diagonal elements of the one-body density matrix) but will not be important in this article. The irreducible diagram corresponding to Eq.~(\ref{eqn:3-cluster}) for $\vectx'=\vectx$ and with $\vectx_3$ traced out is depicted in Fig.\@ \ref{fig:diagrams}(b). In practice it is convenient to omit one-particle irreducible diagrams while stating the particle number of the reduced diagrams explicitly.

\begin{figure}
	\includegraphics[width=.8\linewidth]{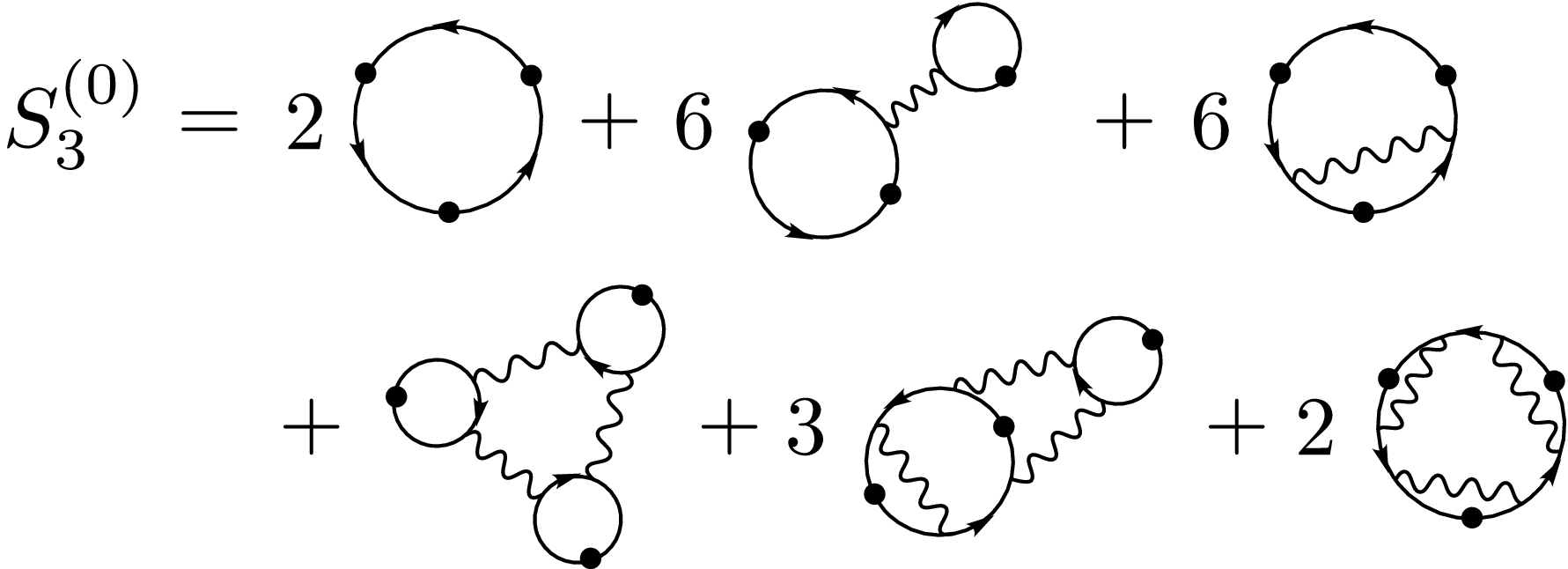}
	\caption{The sum of irreducible three-particle diagrams $S_3^{(0)}$. Most of the diagrams appear with multiplicities larger than 1. The first line includes only the interaction contributions up to order 2, while the second line accounts for all six possible permutations of the third-order interaction contribution. For distinguishable particles only the fourth diagram contributes.}
	\label{fig:cluster_sums}
\end{figure}

Let us now focus on diagrams that appear in the full trace of the cluster expansion, \ie, the partition function, with the purpose of counting only distinct diagrams, then provided with multiplicities. Consider a full diagram in the expansion that is built out of $l$ irreducible diagrams of sizes $n_1\geq \dots \geq n_l$. By distributing the particle indices among the irreducible diagrams in a different way one finds equivalent full diagrams. Therefore, the multiplicity of any such diagram contains the combinatorial factor
\begin{equation}
	\#^N_\mathfrak{N}=\frac{1}{\prod_{\nu=1}^{\infty}m_\mathfrak{N}(\nu)!}\frac{N!}{\prod_{i=1}^{l}n_i!},
	\label{eqn:multinom}
\end{equation}
where $m_\mathfrak{N}(\nu)$ is the multiplicity of the number $\nu$ in $\mathfrak{N}=\{n_1,\dots,n_l\}$. It is the number of possible partitions of the $N$ particle indices into sets of the sizes $n_1, \dots ,n_l$. This holds irrespective of the structure of the irreducible diagrams, whereas an additional factor counts the number of ways to relabel the coordinates inside an irreducible diagram depending on its structure. If we collect all full diagrams in the cluster expansion that factorize into irreducible diagrams of the sizes $n_1,\dots,n_l$ their sum can be written as
\begin{equation}
	\left(\#^N_\mathfrak{N}\right)\prod_{i=1}^{l}S_{n_i}^{(0)},
	\label{eqn:factorization}
\end{equation}
where $S_n^{(0)}$ is the sum of all $n$-particle irreducible diagrams, including the multiplicities from internal relabeling. As an example consider the cluster expansion for three bosons. There are three different partitions of the particles with combinatorial factors
\begin{equation}
	\#_{\{1,1,1\}}^3=1,\qquad \#_{\{2,1\}}^3=3, \qquad \#_{\{3\}}^3=1
\end{equation}
so that the partition function is given by
\begin{equation}
	Z=\left(S_1^{(0)}\right)^3+3 S_2^{(0)}S_1^{(0)}+S_3^{(0)}.
\end{equation}
The sum of the three-particle irreducible diagrams $S_3^{(0)}$ with their individual multiplicities is shown in Fig.~\ref{fig:cluster_sums}.
Let us focus on the multiplicity of the second diagram. It is built from one interaction contribution of order two and one free propagation (order one). Choosing the interacting pair out of three particles already gives three possibilities corresponding to the Ursell decomposition, cf.~Eq.~(\ref{eqn:ursell}). Then, one of the interacting particles has to be linked to the free particle by a permutation. This can be achieved with two distinct exchange permutations, yielding the overall multiplicity of 6. A detailed description of how the coefficients are determined in general can be found in \cite{Hummel2018}.

When dealing with \textit{partial} traces the combinatorial factors in Eq.~(\ref{eqn:multinom}) as well as the multiplicities of the irreducible diagrams with fixed coordinates have to be modified but the general statement remains.

\subsection{The nonlocal pair correlation in cluster expansion}
We focus on the normalized nonlocal pair-correlation function for bosons which, for a homogeneous system with fixed particle number $N$, is defined as
\begin{equation}
g_2^{(N)}(\vect{r})=\frac{\braket{\hat{\Psi}^\dagger(0)\hat{\Psi}^\dagger(\vect{r})\hat{\Psi}(\vect{r})\hat{\Psi}(0)}}{\rho^2},
\label{eqn:g2general}
\end{equation}
where $\hat{\Psi}(\vectx)$ and $\hat{\Psi}^\dagger(\vectx)$ are the bosonic field operators at position $\vectx$ and $\rho=N/V$ is the particle density. By taking the expectation value in the canonical ensemble we can write Eq.~(\ref{eqn:g2general}) in terms of the many-body propagator
\begin{multline}
g^{(N)}_2(\vect{r})= \frac{1}{\rho^2}\frac{N(N-1)}{Z^{(N)}_+}\\
\quad\times\int\Diff{N-2}{\vectx}\K{N}_+(\vectx,\vectx;t=-\i\hbar\beta)\big\vert_{\vectx_1=0,\vectx_2=\vect{r}}
\label{eqn:g2}
\end{multline}
\vspace{0.5cm}
with the canonical partition function
\begin{equation}
Z^{(N)}_+=\int\Diff{N}{\vectx}\K{N}_+(\vectx,\vectx;t=-\i\hbar\beta).
\label{eqn:Z}
\end{equation}
A derivation can be found in Appendix~\ref{app:canonical}.
In the presence of external potentials the normalization has to be replaced according to $\rho^2\mapsto \rho(0)\rho(\vect{r})$. Both numerator and denominator in Eq.~(\ref{eqn:g2}) can now be expanded in terms of cluster diagrams.
Let us define $S_n^{(k)}(\vectx_1',\dots,\vectx'_k|\vectx_1,\dots,\vectx_k)$ as the sum of all $n$-particle irreducible diagrams (including multiplicities from internal relabeling) that have all but $k$ coordinates traced out. It is usually more convenient to work with the rescaled cluster sums
\begin{equation}
	B_n^{(k)}\equiv\frac{S_n^{(k)}}{(n-k-\delta_{k0})!}
	\label{eqn:B_def}
\end{equation}
where the factorial accounts for the multiplicity of the cycle diagrams in the noninteracting case and $\delta_{k0}$ is the Kronecker $\delta$. The functions $B_n^{(k)}$ are recursively related to each other by
\begin{multline}
\int\diff{\vectx_{k+1}}B_n^{(k+1)}(\vectx_1',\dots,\vectx'_k,\vectx_{k+1}|\vectx_1,\dots,\vectx_k,\vectx_{k+1})\\
=\frac{(n-k-\delta_{k0})!}{(n-k-1)!}B_n^{(k)}(\vectx_1',\dots,\vectx'_k|\vectx_1,\dots,\vectx_k).
\label{eqn:Bn_reduction}
\end{multline}
With these definitions we can partially factorize the cluster expansion similar to Eq.~(\ref{eqn:factorization}), but we have to distinguish the case where the fixed coordinates $\vectx_1$ and $\vectx_2$ belong to the same irreducible diagram from the case where they belong to different ones. Making use of the cluster expansion of the partition function leads to the general result
	\begin{multline}
	g_2^{(N)}(\vect{r})=\frac{1}{\rho^2 Z^{(N)}}
	\left\{
	\sum_{k=2}^N B_k^{(2)}(0,\vect{r})Z^{(N-k)}\right.
	\\
	\left. +\sum_{k=1}^{N-1}\sum_{l=1}^{N-k}B_k^{(1)}(0)B_l^{(1)}(\vect{r})Z^{(N-k-l)}
	\right\},
	\label{eqn:g2cluster_expansion}
	\end{multline}
where we use the shorthand notation $B_k^{(2)}(\vectx,\vect{y})\equiv B_k^{(2)}(\vectx,\vect{y}|\vectx,\vect{y})$ and $B_k^{(1)}(\vectx)\equiv B_k^{(1)}(\vectx|\vectx)$ and omitted the index $+$ in the partition functions as this (purely combinatorial) result is not restricted to bosons. The partition function can be conveniently calculated from the recursion relation
\begin{equation}
Z^{(N)}=\frac{1}{N}\sum_{k=1}^N B_k^{(0)} Z^{(N-k)}
\label{eqn:Zrecursion}
\end{equation}
that stems from purely combinatorial calculations, too. In both Eqs.~(\ref{eqn:g2cluster_expansion}) and (\ref{eqn:Zrecursion}) we have defined $Z^{(0)}=1$.

\subsection{Short-time approximation}
\label{sec:short-time}
Up to this point the expression for $g_2^{(N)}$ is exact but purely formal. A key step now is to realize that for temperatures above the quantum degenerate regime it is sufficient to include short-time information on the propagators $\K{n}$ to be specified in the following. For finite systems without external potentials we replace all the propagators in the calculation by their infinite space equivalents, \ie, we assume that the particles do not explore the whole system in arbitrarily short times. The condition for this approximation to be accurate can be estimated at the single-particle level to be
\begin{equation}
	t\ll \frac{mV^{\frac{2}{D}}}{2\pi\hbar}\equiv t_\mathrm{T},
	\label{eqn:time_condition}
\end{equation}
where $m$ is the mass of the particle, $V$ is the volume of the system, and $D$ is the dimension. The characteristic time $t_\mathrm{T}$ can be thought of as the typical traversal time through the system of a particle with momentum $\hbar V^{-1/D}$, where the latter corresponds to the minimal uncertainty in the momentum of a wave packet in the volume $V$. If we switch to imaginary time the condition (\ref{eqn:time_condition}) can be translated into
\begin{equation}
	\lambda_T^D\ll V,
\end{equation}
introducing the thermal (de Broglie) wavelength
\begin{equation}
	\lambda_T=\sqrt{\frac{2\pi\hbar^2\beta}{m}}.
\end{equation}
At this length scale the propagator of a single free particle decays in imaginary time $t=-\i\hbar\beta$. This gives the intuitive picture that within the regime of validity of the short-time approximation all clusters of particles have a characteristic size that scales with $\lambda_T$ that is much smaller than any length scale introduced from external confinement, such that their internal structure is essentially independent of the latter. We have to stress that the short-time approximation does not require that the thermal wavelength be small compared to the mean interparticle separation, \ie, we can have
\begin{equation}
	\frac{N\lambda_T^D}{V}> 1,
\end{equation}
in contrast to the case of high-temperature expansions in the thermodynamic limit.

In the presence of smooth external potentials the short-time approximation can be modified such that only the internal dynamics of a cluster is mapped to infinite space, while its center of mass evolves according to the single-particle (short-time) propagator \cite{hummel2016}. 

The short-time approximation defined above is well known in semiclassical physics, where it corresponds to taking into account only the shortest classical paths in the Van Vleck-Gutzwiller propagator \cite{Gutzwiller1990}. It can thus be easily extended to include, \eg, corrections from boundaries, which has its direct application in the calculation of the mean density of states, known as Weyl's law \cite{hummel2013}. The short-time approximation of the propagator thereby encodes the information on the slowly varying parts of the density of states. Note that the bound in Eq.~(\ref{eqn:time_condition}) plays the role of a Heisenberg time $t_\mathrm{H}=2\pi\hbar/\Delta$ where $\Delta$ is the mean single-particle level spacing. It can be understood as a lower bound for the time needed to resolve the discreteness of the spectrum. This means that the price we pay for using the short-time approximation is the loss of all information related to this discreteness.

The power of the short-time approximation lies in the high level of generality leading to certain general scaling properties. We first focus on the full trace of a cluster as it appears, \eg, in the partition function. For homogeneous systems the short-time approximation tells us that, due to translational invariance, every cluster contributes with a factor proportional to the volume of the system (the presence of smooth external potentials results in an effective volume \cite{hummel2016}). For $D$-dimensional homogeneous systems this will lead to a volume factor $V$ for every fully traced cluster. Now we assume an interaction potential $U$ that depends only on the coordinates $\vectx$, an interaction parameter $\alpha$ with the dimension of energy, and the physical constants $m$ and $\hbar$. A dimensional analysis then shows that we can write the potential as $\alpha\tilde{U}(\sqrt{\bar{\alpha}}\vectx/\lambda_T)$ in terms of a dimensionless function $\tilde{U}(\vect{y})$, a dimensionless parameter $\bar{\alpha}=\beta\alpha$, and $\lambda_T$.
Using this scale transformation we can rewrite the interaction contributions
\begin{equation}
	\dK{n}(\vectx',\vectx;t=-\i\hbar\beta)=\lambda_T^{-nD}\Delta\tilde{K}^{(n)}\left(\frac{\vectx'}{\lambda_T},\frac{\vectx}{\lambda_T};\bar{\alpha}\right)
\end{equation}
as a dimensionless function $\Delta\tilde{K}^{(n)}$. This implies very generally that in the short-time approximation the functions $B_n^{(k)}$ in Eqs.~(\ref{eqn:B_def})--(\ref{eqn:Zrecursion}) will be proportional to $\lambda_T^{-kD}$ for $k>0$ or to $V/\lambda_T^D$ for $k=0$. To make this explicit we define the dimensionless functions
\begin{align}
b_n^{(k)}&=\lambda_T^{kD} \,B_n^{(k)}\qquad \text{for }k>0,\nonumber\\
b_n^{(0)}&=\frac{\lambda_T^D}{V}B_n^{(0)},
\label{eqn:rescaled_cluster_sums_general}
\end{align}
that only depend on rescaled variables such as $\vectx/\lambda_T$ and $\bar{\alpha}$. A direct implication is that the nonlocal pair-correlation function $g_2^{(N)}(\vect{r})$ can be written as a rational function in the parameter $V/\lambda_T^D$ with coefficients that depend only on the functions $b_n^{(k)}$ (with $k=0,1,2$). The partition function takes the form of a polynomial in $V/\lambda_T^D$ with coefficients $b_n^{(0)}$, whereas the factor $\rho^{-2}\propto V^2$ compensates for the missing volume dependence in the numerator of the cluster expansion (\ref{eqn:g2cluster_expansion}) for $g_2^{(N)}$.

\section{Application to Lieb-Liniger gas}
\label{sec:LL}
\subsection{The model}
	We now apply the methods of the previous section to compute the pair correlation for the case of $N$ bosons with repulsive short-range interactions in a 1D ring geometry. We describe this system by the well-known Lieb-Liniger (LL) model defined by the Hamiltonian \cite{lieb1963,korepin1997}
	\begin{equation}
		\hamilton=\frac{\hbar^2}{2m}\left(\sum_{i=1}^{N}-\secondderivative{x_i}+c\sum_{\underset{i\neq j}{i,j=1}}^{N}\delta(x_i-x_j)\right)
		\label{eqn:LL-hamiltonian}
	\end{equation}
with $c\geq 0$, $x_i\in [-L/2,L/2]$, where $L$ is the system size, and periodic boundary conditions. The relevant dimensionless coupling parameter in the weakly degenerate regime is $c\lambda_T$. The symmetric eigenfunctions of the Hamiltonian (\ref{eqn:LL-hamiltonian}) can be found via a Bethe ansatz, where periodicity leads to a quantization condition in terms of $N$ coupled transcendental equations \cite{lieb1963}.

In the limit $L\to\infty$, sometimes referred to as extended LL model, the spectrum becomes continuous. The symmetrized many-body propagator for this extended system is known exactly from integrating over all Bethe ansatz solutions \cite{Tracy2008,Dotsenko2010,Prolhac2011}. We were able to rederive this propagator using the closed-form expressions for the wave functions introduced in \cite{korepin1997} to get the strikingly simple form
\begin{equation}
	\K{N}_+(\vectx',\vectx;t)=\frac{1}{N!}\sum_{P\in S_N}\bar{K}^{(N)}(P\vectx',\vectx;t)
	\label{eqn:LL-propagator}
\end{equation}
with
\begin{align}
	\bar{K}^{(N)}(\vectx',\vectx;t)&=\frac{1}{(2\pi)^N}\int\Diff{N}{k}\eh{-\frac{\i\hbar t}{2m}\vectk^2+i\vectk(\vectx'-\vectx)}\nonumber\\
	&\quad\times \prod_{j>l}\frac{k_j-k_l-ic \sgn(x_j'-x_l')}{k_j-k_l-ic\sgn(x_j-x_l)}.
	\label{eqn:LL-eff_propagator}
\end{align}
A derivation of this result can be found in Appendix~\ref{app:propagator_derivation}.
Note that the function $\bar{K}$ is \textit{not} the many-body propagator for distinguishable particles but can be used as a substitute in the cluster expansion for bosons. Since only symmetry-projected quantities matter eventually we are free to replace the interaction contributions $\dK{n}$ in the Ursell decomposition (\ref{eqn:cluster_expansion}) by their symmetry-projected equivalents $\dK{n}_+$. The corresponding expressions for $n=2,3$ can be found in Appendix~\ref{app:dK3}. The nonsymmetrized expression for $\dK{2}$ can be calculated from the propagator for a $\delta$ potential directly, which gives exactly the same result, as it is already symmetric with respect to particle exchange (antisymmetric states are not affected by the $\delta$ potential). The corresponding derivation can be found in Appendix~\ref{app:dK3}.

\subsection{Lieb-Liniger model for three particles---Full cluster expansion}
We will first address the full cluster expansion for $N=3$ particles calculated from the propagator (\ref{eqn:LL-eff_propagator}). As discussed in Sec.~\ref{sec:short-time}, it comes as a rational function in $L/\lambda_T$ with coefficients that are dimensionless functions of the rescaled quantities $r/\lambda_T$ and $c\lambda_T$. Due to the homogeneity of the system the diagonal part of $b_n^{(1)}$ does not depend on $r$, leading to the identification
\begin{equation}
b_n^{(1)}(r)=b_n^{(0)}\equiv b_n.
\end{equation}
By expanding the general result for $g_2^{(N)}$, Eq.~(\ref{eqn:g2cluster_expansion}), with the help of Eq.~(\ref{eqn:Zrecursion}) for $N=3$ and using the shorthand notation $b_n^{(2)}(r)=b_n^{(2)}(0,r)$ we can write the nonlocal pair-correlation function as
\begin{equation}
	g^{(3)}_2(r)=\frac{2}{3}\times\frac{1+[b_2^{(2)}(r)+2b_2\frac{\lambda_T}{L}]+b_3^{(2)}(r)\frac{\lambda_T}{L}}{1+3b_2\frac{\lambda_T}{L}+2b_3\left(\frac{\lambda_T}{L}\right)^2}.
	\label{eqn:g2N3}
\end{equation}
We have calculated the functions $b_n^{(2)}(r)$ and $b_n$ for $n=2,3$ from the interaction contributions $\dK{2}_+$ and $\dK{3}_+$. For $n=2$ we get the simple result
\begin{align}
	b_2^{(2)}(r)&=\eh{-\rt^2}\left[1-\sqrt{4\pi}\ct\,\eh{(\ct+|\rt|)^2}\erfc(\ct+|\rt|)\right],\\
	b_2&=\frac{1}{\sqrt{2}}\left[2\eh{\ct^2}\erfc(\ct)-1\right],\label{eqn:b2}
\end{align}
where $\rt=\sqrt{2\pi} r/\lambda_T$ is the distance in terms of the thermal wavelength and
\begin{equation}
	\ct=\lambda_T c/\sqrt{8\pi}
	\label{eqn:ct}
\end{equation}
is the dimensionless (thermal) interaction strength. The corresponding expressions for $n=3$ are more complicated and can be found in Appendix~\ref{app:dK3}, Eqs.~(\ref{eqn:d1})--(\ref{eqn:d3}), and Eq.~(\ref{eqn:b3}). The integrated function $b_2$ in Eq.~(\ref{eqn:b2}) is closely related to the virial coefficient found in \cite{Hoffman2015} for the spin-balanced Gaudin-Yang model. The correct normalization $\int\diff{r}g_2^{(3)}(r)=2L/3$ is obtained from Eq.~(\ref{eqn:Bn_reduction}) only if the integration domain $(-L/2,L/2)$ can be replaced by $\mathbb{R}$ in all nontrivial integrals in the spirit of the short-time approximation, \ie, if $b_{2,3}^{(2)}(r)\approx 0$ for $|r|>L/2$. In the case at hand this gives the natural bound $\lambda_T\lesssim L/2$ for the short-time approximation to be valid as both $b_{2,3}^{(2)}$ have a typical extent of $\lambda_T$. This means that we can make predictions for very low temperatures as long as the semiclassical result for $g_2^{(N)}(r)$ saturates well before $r=L/2$.

For comparison with numerical results we calculated the exact correlation function using the Bethe ansatz solutions similar to \cite{Zill2016}. The details can be found in Appendix~\ref{app:numerics}. It is straightforward to show that the system size $L$ can be eliminated completely from $g_2$ in both results using the scale transformation $x_i\mapsto x_i/L$, $k_i\mapsto k_iL$, $c\mapsto cL$, $\beta\mapsto \beta/L^2$, where the $k_i$ are the quasimomenta that appear in the Bethe solutions. We thus express $r$ and $\lambda_T$ in units of $L$ in all plots and use $\lambda_T$ as the temperature parameter rather than $T$ or $\beta$.
\begin{figure}
	\includegraphics[width=\linewidth]{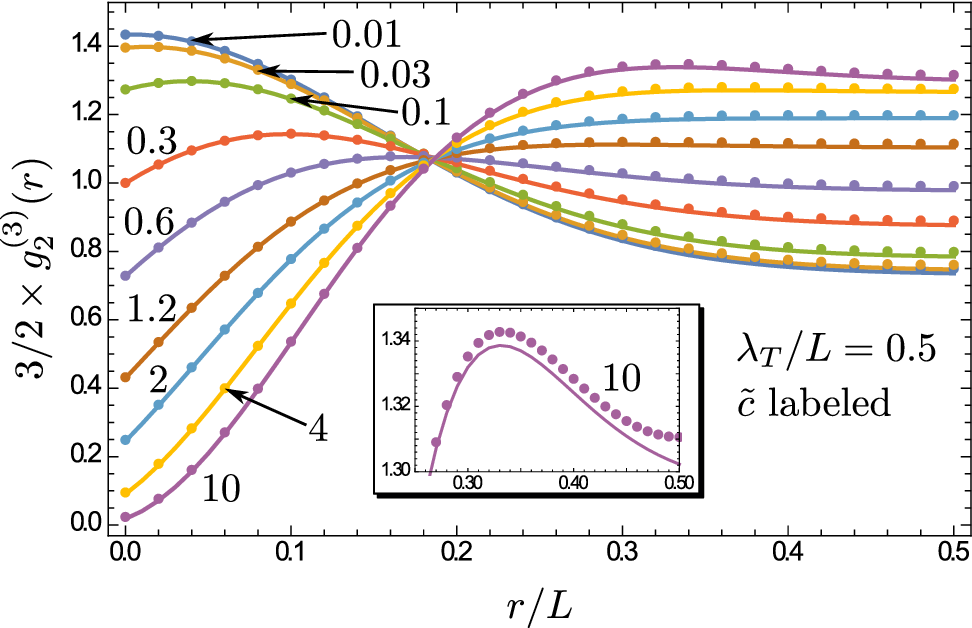}
	\caption{Comparison of $g_2^{(3)}(r)$, Eq.~(\ref{eqn:g2N3}), (solid lines) with numerical calculations (dots) for $\lambda_T/L=0.5$ and various values of the thermal interaction strength $\ct$, Eq.~(\ref{eqn:ct}) (labeled). The inset shows the maximum arising for $\ct=10$, an indicator of quasicrystalline order.}
	\label{fig:N3lambda05}
\end{figure}
Figure \ref{fig:N3lambda05} shows $3/2 g_2^{(3)}(r)$ for various values of $\ct$, Eq.~(\ref{eqn:ct}), and for $\lambda_T/L=0.5$. The absolute and relative error in the semiclassical results are smaller than $10^{-2}$ for all values of $\ct$ at this temperature. For higher temperatures the results are more accurate, \eg, for $\lambda_T/L=0.3$ (not shown) both the absolute and relative error of the semiclassical result are of the order $10^{-6}$ for all values of $\ct$.
Considering the fact that for $\lambda_T/L=0.5$ the numerical calculations converge up to an error of 0.1\% already for a summation cutoff after only 15--30 states (depending on the interaction strength), the accuracy of the semiclassical prediction based on a continuous spectrum is impressive.

Interestingly, a feature that usually becomes visible only for very low temperatures, the nonmonotony of $g_2^{(N)}$ in the fermionization regime of large $\ct$ \cite{Deuar2009,Cherny2006}, can already be seen in Fig.\@ \ref{fig:N3lambda05}. There, the maximum value of $g_2^{(3)}(r)$ at $r/L\approx1/3$ for $\ct=10$ is highlighted in the inset and can be interpreted as a precursor of a quasicrystalline order in the two-particle correlations. For larger values of $\lambda_T>0.5L$ the approximation fails as expected.

\subsection{Exploiting the universal scaling of the short-time approximation}
\begin{figure}
	\includegraphics[width=\linewidth]{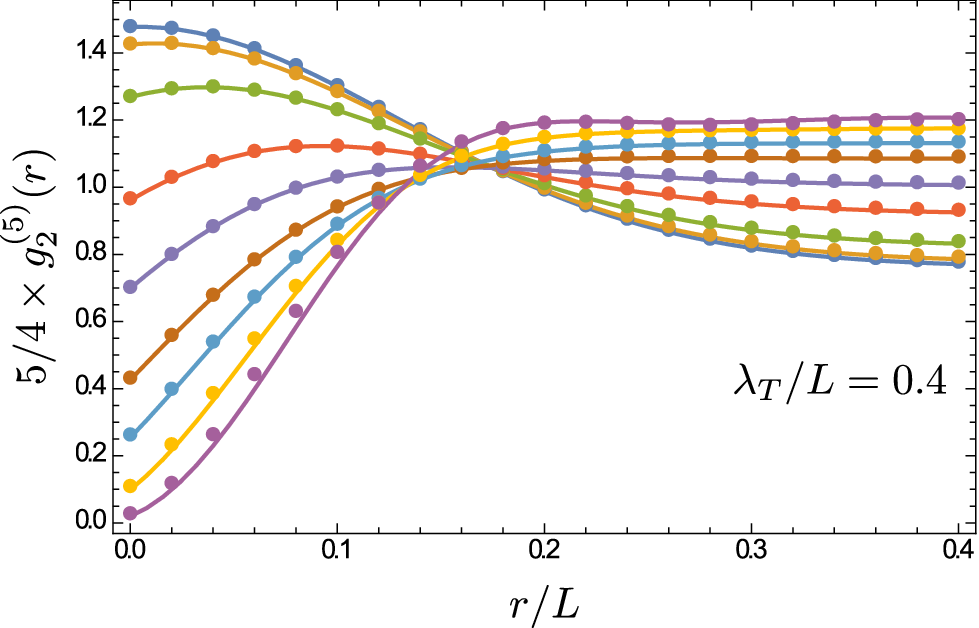}
	\caption{The nonlocal pair-correlation function for $N=5$ particles from Bethe ansatz calculations (dots) and the semiclassical result (solid lines) for $\lambda_T/L=0.4$ using the functions $b_n^{(2)}(r)$ and $b_n^{(0)}$ Eq.~(\ref{eqn:rescaled_cluster_sums_general}) for $n=4,5$ that have been recursively extracted from the numerical results for $g_2^{(4)}(r)$ and $g_2^{(5)}(r)$ at $\lambda_T/L=0.1$. The values for $\ct$ (ranging from 0.01 to 10, from top to bottom at $r=0$) are the same as in Fig.~\ref{fig:N3lambda05}.}
	\label{fig:g2N5lmbd04}
\end{figure}
The general scaling properties of the short-time approximation that we found in Sec.~\ref{sec:short-time} are not only useful to identify relevant parameters of the theory but can actually be used as a predictive tool. Let us assume that we know the expressions for $b_n$ and $b_n^{(2)}(r)$ up to a certain cluster size $n=l-1$. If we can find, in whatsoever way, \eg, by direct measurement \cite{Schweigler2017}, an expression for $g_2^{(l)}(r)$ for fixed values of $\ct$ and (small enough) $\lambda_T$ it contains all the information we need to calculate $b_{l}$ and $b_{l}^{(2)}(r)$. The scaling behavior of the latter can then be used to calculate $g_2^{(l)}(r)$ at all temperatures in the range of validity of the short-time approximation with the same $\ct$ or to find better approximations for higher particle numbers (see next section). The interplay between the scaling of the functions $b_n^{(k)}$ and the form of $g_2^{(N)}$ as a rational function in $\lambda_T/L$ renders this approach nontrivial. To actually calculate $b_l^{(2)}$ and $b_l$ from $g_2^{(l)}$, we note that $b_n^{(2)}(r)\to 0$ for $r\to\infty$ and that the cluster expansion of $g_2^{(l)}$ contains $b_l$ only in the denominator. This means that $g_2^{(l)}(r)/g_2^{(l)}(\infty)$ depends on $b_l^{(2)}(r)$ but not on $b_l$, while the latter can be found independently from $g_2^{(l)}(\infty)$. In practice, the diverging argument $r\to\infty$ has to be replaced by a value that lies inside the saturation regime of $g_2^{(l)}$. This explains why we have to know $g_2^{(l)}(r)$ for ``small'' values of $\lambda_T$.
As the above considerations use only the homogeneity of the system, they are not restricted to 1D or to $\delta$-like interaction potentials.

To demonstrate the power of the method, we have used the numerical results from the Bethe ansatz calculations of $g_2^{(4)}(r)$ and $g_2^{(5)}(r)$ at $\lambda_T=0.1L$ and various values of $\ct$ to calculate the clusters $b_n^{(2)}$ and $b_n^{(0)}$ for $n=4,5$. The results have then been used to calculate $g_2^{(5)}(r)$ at $\lambda_T/L=0.4$. The comparison of the respective predictions with the numerical calculations is shown in Fig.~\ref{fig:g2N5lmbd04}.
The nearly perfect agreement for all values of the interaction strength shows that the method is indeed applicable to the case at hand. 

We have investigated the breakdown of the validity of our approach by calculating the mean absolute error in the semiclassical results for $g_2^{(N)}(r)$ using the 2-norm
\begin{equation}
	\overline{\Delta g_2^{(N)}}=\sqrt{\frac{1}{\lambda_T}\int_{0}^{\lambda_T}\diff{r}\left[\Delta g_2^{(N)}(r)\right]^2},
\end{equation}
where $\Delta g_2^{(N)}(r)$ is the difference between the numerical and semiclassical results.
\begin{figure}
	\includegraphics[width=\linewidth]{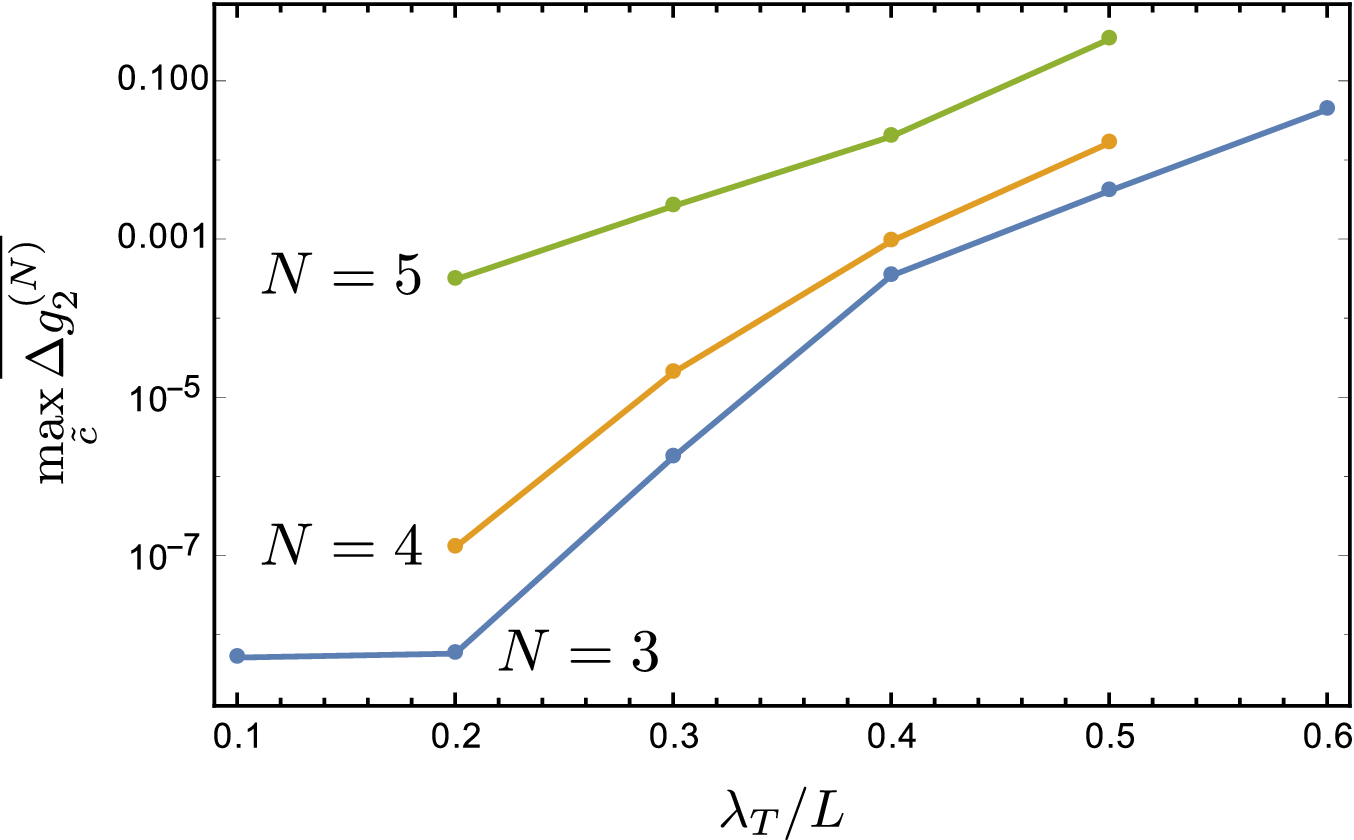}
	\caption{The maximum (with respect to the interaction strength $\ct$) of the mean difference between semiclassical and numerical results (see text). While the deviation is smaller than the numerical precision for $N=3,\lambda_T\leq 0.2$ it increases rapidly for $\lambda_T/L\geq0.2$.}
	\label{fig:error}
\end{figure}
Figure \ref{fig:error} shows the maximum of this mean error with respect to the interaction strength ranging from $0.01$ to $10$ for $N=3,4,5$ and for various values of $\lambda_T/L$. For $N=3$ and $\lambda_T\leq0.2$ the error is smaller than the numerical precision (see Appendix~\ref{app:numerics}). The deviation for $\lambda_T=0.1L$ is not shown for $N=4,5$, as this is the value used for the extraction of the functions $b_n^{(0)},b_n^{(2)}(r)$ for $n=4,5$. The large offset between the graphs for the different particle numbers can be explained by the rather small numerical precision in the extracted cluster contributions, but all three curves show a roughly exponential increase in the range of $0.1\leq\lambda_T/L\leq 0.5$, indicating a sudden breakdown of the short-time approximation.
\subsection{Truncated cluster expansion for higher particle numbers}
The full cluster expansion for $g_2$, in principle, could be calculated from the propagator (\ref{eqn:LL-eff_propagator}) for arbitrary particle numbers $N$. In practice one would have to (partially) trace not only $\dK{n}$ for $1\leq n\leq N$, which is a difficult task, but also all permutations of different products thereof. Here we will use only the information from interaction contributions up to third order. One way to achieve this goal is to truncate the expansion into interaction contributions, Eq.~(\ref{eqn:cluster_expansion}), to take into account only the desired orders. This has been proven to yield excellent results for the canonical partition function with a truncation to second-order interaction contributions \cite{hummel2016}. The resulting expressions comprise clusters of all sizes due to the symmetrization of the propagator. But already at the level of cluster sizes $n\leq3$ we can make good predictions for certain regimes while using only such minimal information. As argued above, the full cluster expansion is a rational function in the parameter $\lambda_T/L$ with coefficients that are functions of $\rt$, $\ct$, and $N$. With $A_n(r)=b_n^{(2)}(r)-(n-1)b_n\lambda_T/L$ we can write $g_2$ as
\begin{widetext}
	\begin{equation}
	g_2^{(N)}(r)=\frac{N-1}{N}\left\{
	1+\frac{
		A_2(r)+\left[\binom{N-2}{1}A_3(r)+\binom{N-2}{2}b_2A_2(r)\right]\frac{\lambda_T}{L}+\mathcal{O}(2)
	}{
		1+\binom{N}{2}b_2\frac{\lambda_T}{L}+\mathcal{O}(2)
	}
	\right\},
	\label{eqn:truncated_cluster_expansion}
	\end{equation}
where $\mathcal{O}(2)$ stands for higher orders in $\lambda_T/L$. We can now expand this function into a formal series in the parameter $\lambda_T/L$, while treating the functions $A_n$ as constants to preserve normalization. This results in
\begin{equation}
g_2^{(N)}(r)\approx\frac{N-1}{N}\bigg\{1+A_2(r)
+\Big[(N-2)A_3(r)-(2N-3)A_2(r)b_2\Big]\frac{\lambda_T}{L}\bigg\}.
\label{eqn:virial_expansion}
\end{equation}
\end{widetext}
The terms of order $n$ in $\lambda_T/L$ now come with a polynomial in the particle number $N$ that is of the order $n$, a fact that is well hidden in the rational expression for $g_2^{(N)}$. The series expansion has a positive convergence radius for any finite particle number, and the truncation is a good approximation if we take the ratio between the thermal wavelength and the mean interparticle distance,
\begin{equation}
	n_T=N\lambda_T/L,
	\label{eqn:nt}
\end{equation}
as a small parameter. 
\begin{figure}
	\includegraphics[width=\linewidth]{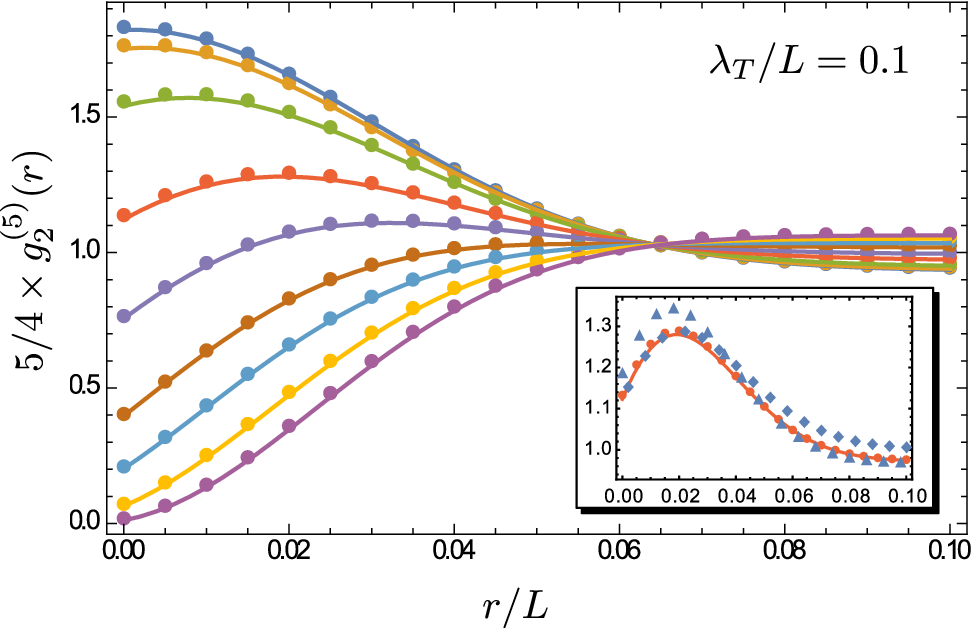}
	\caption{Comparison of the expansion for $g_2^{(5)}$, Eq.~(\ref{eqn:virial_expansion}), with numerical results (dots) for $\lambda_T/L=0.1$ for the same range of values of $\ct$ as in previous figures. The inset shows the effect of truncating the expansion of $g_2^{(5)}$ after the single two-particle clusters [first line in Eq.~(\ref{eqn:virial_expansion})] (triangles) and the effect of neglecting all coefficients that are subleading in the particle number (squares), \ie, Eq.~(\ref{eqn:thermodynamic_limit}), for $\ct=0.3$.}
	\label{fig:g2N5lmbd01virialvsnum}
\end{figure}
Figure \ref{fig:g2N5lmbd01virialvsnum} shows the comparison of Eq.~(\ref{eqn:virial_expansion}) with numerical calculations for $N=5$ particles and $N\lambda_T/L=0.5$. The agreement is very good for the whole range of interaction parameters $\ct$. The inset shows the effect of truncating the expansion Eq.~(\ref{eqn:virial_expansion}) to single two-particle clusters (first two terms in the equation) and the effect of neglecting terms of subleading order in the particle number, respectively, for $\ct=0.3$ (the latter corresponds to the thermodynamic limit that will be addressed below). Clearly, there is a major improvement by using the additional information from $b_3^{(0)},b_3^{(2)}(r)$, and multiple clusters, where finite-size effects play a crucial role. Also note that the fermionic limit $\ct\to\infty$ at $r=0$ yields zero for all orders in the full expansion, which is often referred to as antibunching. Thus, in this limit the error in the truncated expansion (\ref{eqn:virial_expansion}) is of the order $n_T^2/N$.
\subsection{The thermodynamic limit}
\begin{figure}
	\includegraphics[width=\linewidth]{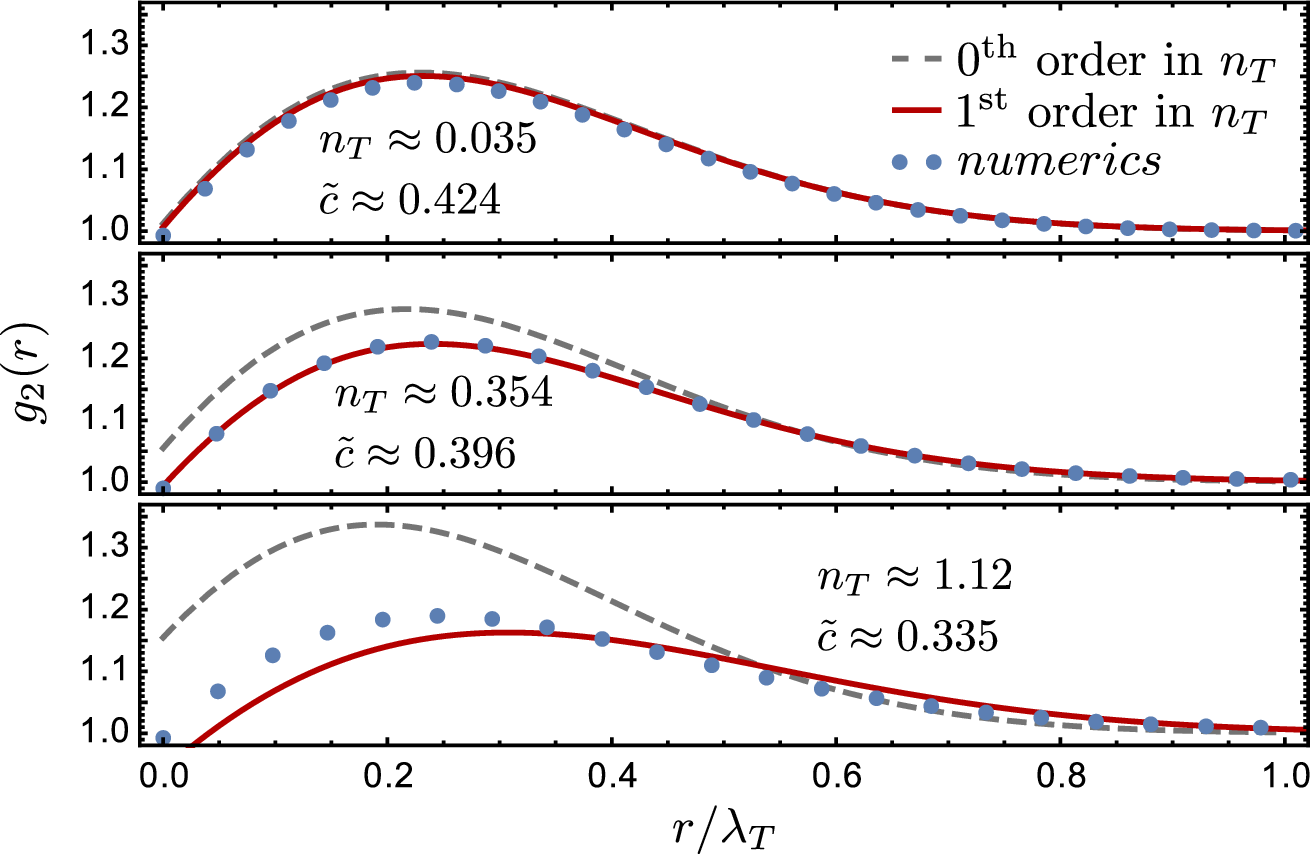}
	\caption{Nonlocal pair correlation in the thermodynamic limit for different interaction strengths and different values of $n_T$ such that $g_2(0)\approx 1$. In the high-temperature or low-density regime $n_T\ll 1$ only two-particle clusters contribute. For lower temperatures $\mathcal{O}(n_T)$ corrections cannot be neglected and larger clusters play a role.}
	\label{fig:high_to_low_T_transition}
\end{figure}

From the virial-like expansion Eq.~(\ref{eqn:virial_expansion}) it is easy to find the thermodynamic limit by omitting all terms that are subleading in $N$ while fixing $n_T$, Eq.~(\ref{eqn:nt}). This gives
\begin{equation}
	g_2(r)=1+b_2^{(2)}(r)+[b_3^{(2)}(r)-2b_2b_2^{(2)}(r)]n_T
	+\mathcal{O}(n_T^2).
	\label{eqn:thermodynamic_limit}
\end{equation}
Equation (\ref{eqn:thermodynamic_limit}) can also be found within a grand-canonical approach by inverting the fugacity expansion in terms of the particle number in the high-temperature limit \cite{Hummel2018}. 
\begin{figure}
	\includegraphics[width=\linewidth]{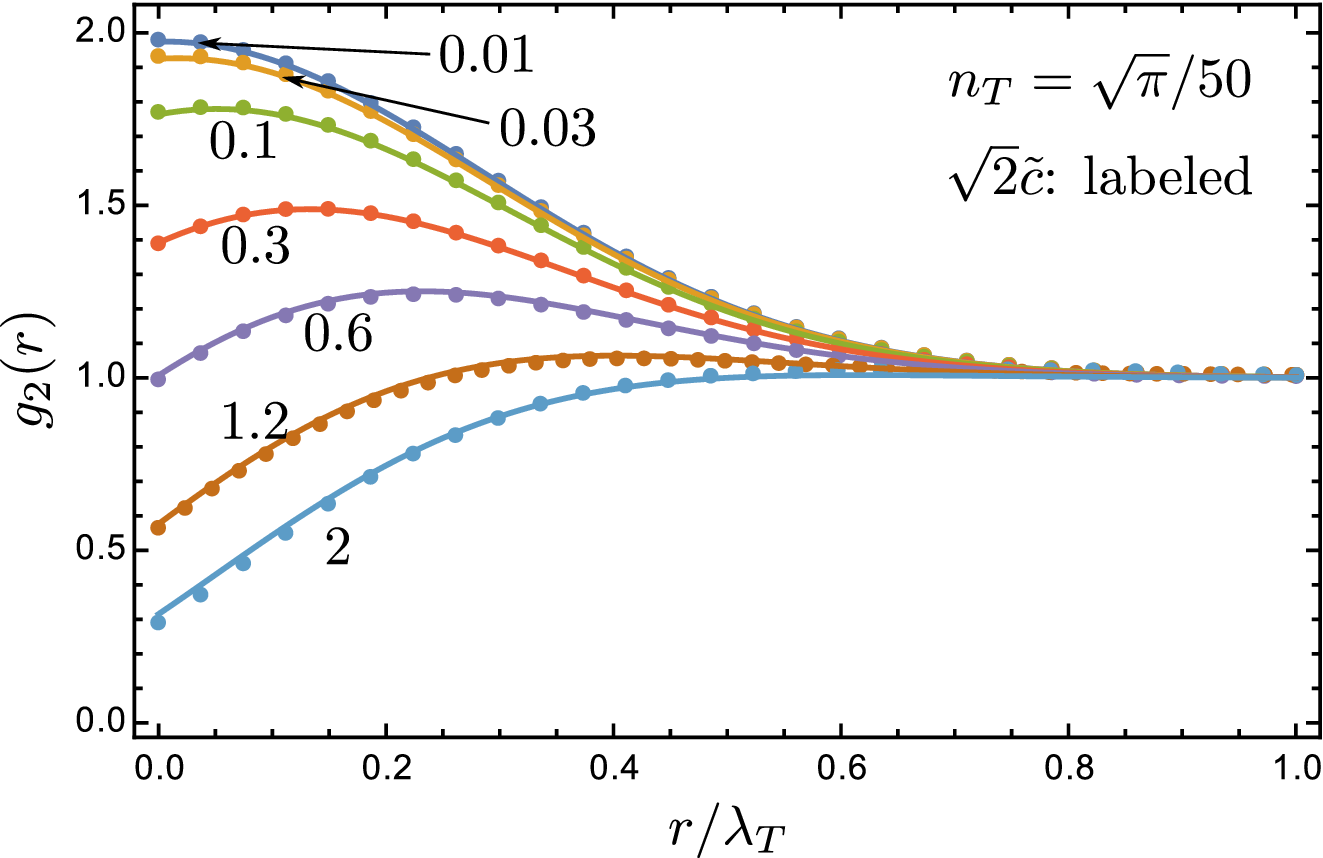}
	\caption{Comparison of numerical results for $g_2(r)$ from \cite{Deuar2009} (for error estimates see \cite{Deuar2009}) with Eq.~(\ref{eqn:thermodynamic_limit}) for $n_T\approx 0.035$ and $\ct$ labeled. The numerical method in \cite{Deuar2009} cannot access the fermionization regime $\ct\gg 1$.}
	\label{fig:thermodynamic_limit}
\end{figure}
A comparison with numerical results obtained in \cite{Deuar2009} is shown in Figs.\@ \ref{fig:thermodynamic_limit} and \ref{fig:high_to_low_T_transition}. Figure~\ref{fig:thermodynamic_limit} demonstrates the validity of our result in the full range of interactions. For high temperatures (low densities) $n_T\ll 1$ it suffices to take into account only $b_2^{(2)}(r)$. For higher values of $n_T$ the next-order term gives non-negligible corrections.
Figure \ref{fig:high_to_low_T_transition} shows $g_2(r)$ for $n_T=\sqrt{\pi/2500},\sqrt{\pi/25},\sqrt{\pi/2.5}$ and $\ct^2=0.18,0.1568,0.1125$, respectively. For $n_T=\sqrt{\pi/2500}$, $g_2$ can be approximated by single- and two-particle clusters. For higher values of $n_T$ the $\mathcal{O}(n_T)$ contributions, and thus three-particle clusters, have to be included, and for $n_T=\sqrt{\pi/2.5}\approx 1.12$ the truncation to first order in $n_T$ is not sufficient anymore for a precise prediction but still gives reasonable qualitative agreement with numerical calculations.
\begin{figure}
	\includegraphics[width=\linewidth]{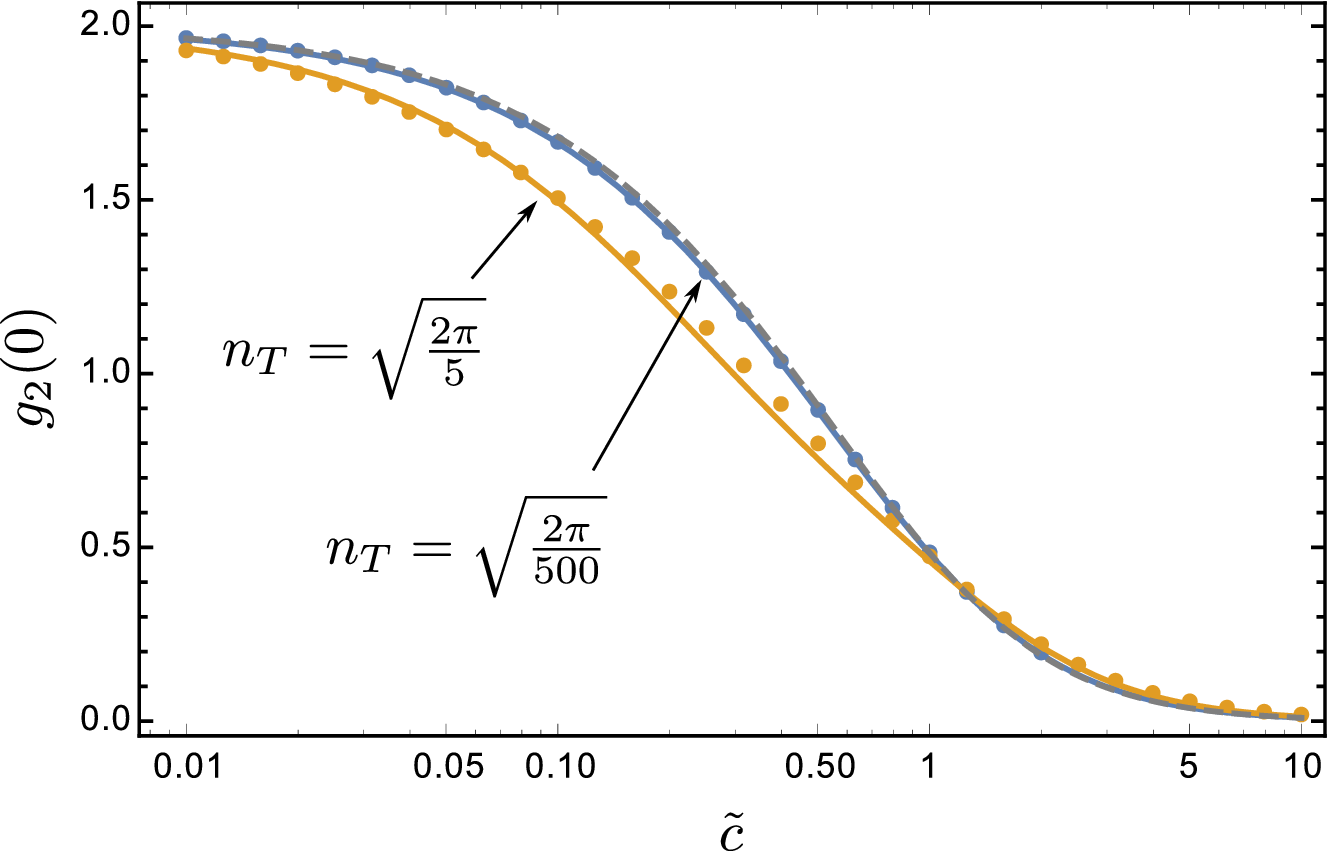}
	\caption{Local correlations $g_2(0)$ with respect to the interaction parameter. Numerical data (dots) is taken from \cite{Kheruntsyan2003}. The approximation by two-particle clusters (gray dashed) is sufficient for high temperatures (low densities). By including the next order in the cluster expansion (solid line) we can see a major improvement in the regime of lower temperature.}
	\label{fig:local_correlations}
\end{figure}
Figure \ref{fig:local_correlations} shows the local correlations $g_2(0)$ for a wide range of the interaction parameter. By including the first-order correction in $n_T$ we can see a major improvement in the agreement of numerical (taken from \cite{Kheruntsyan2003}) and semiclassical results. Note that the local version $g_2(0)$ of the pair correlation can be calculated exactly by solving integral equations using the Hellmann-Feynman theorem \cite{Kheruntsyan2003} (higher local correlation functions have been found from viewing the LL model as a limiting case of the sinh-Gordon model \cite{Kormos2011}), but to the best of our knowledge, all published analytical results for $g_2(r)$ in the weakly degenerate regime were derived in perturbation theory, \ie, they are only valid in the limits of weakly or strongly interacting bosons. Our result, Eq.~(\ref{eqn:thermodynamic_limit}), represents the generalization of these results for arbitrary interaction strengths in the moderate- to high-temperature regime.
\section{Conclusion and summary}
	In this paper we have addressed the spatial structure of few- and many-body states in interacting quantum systems by means of the nonlocal correlation functions. Using a combination of two key ingredients, namely, neglecting the discreteness of the extremely dense many-body spectrum and including interaction effects nonperturbatively by means of cluster expansions, we derived analytical formulas for two-point correlators covering a wide range of temperatures and interaction strengths. 
	
	Specifically, by making use of the method of Ursell operators we developed an exact formula for the nonlocal pair-correlation function and the partition function for finite particle numbers in terms of sums of irreducible (cluster) diagrams. We then showed how, in the high-temperature regime $\lambda_T^D\ll V$, these diagrams can be calculated from semiclassical short-time approximations of the quantum-mechanical many-body propagators. We used these methods to calculate explicit analytical formulas for the nonlocal pair-correlation function in the Lieb-Liniger gas for temperatures above quantum degeneracy that we compared with numerical calculations based on the exact Bethe ansatz solutions of the model. For the example of three particles we showed that the full cluster expansion in short-time approximation remains valid up to $\lambda_T\approx L/2$. We then demonstrated that the universal scaling behavior of the latter remains valid for higher particle numbers. This was done by predicting the form of the nonlocal pair-correlation function for a whole range of temperatures by rescaling the numerical values for a fixed temperature. Comparing the results obtained from this rescaling procedure to numerical calculation showed very good agreement down to $\lambda_T=0.4L$. For higher particle numbers we presented approximations that are valid well above the quantum degeneracy regime, \ie, $N\lambda_T/L\ll1$, while still explicitly depending on the particle number and thus explicitly accounting for its finiteness. Finally, by neglecting the contributions that are subleading in the particle number, we presented the exact results for the first two orders of the series expansion of $g_2(r)$ in the quantum degeneracy parameter $n_T=N\lambda_T/L$ and showed that it agrees well with the numerical results that were obtained by other authors. 
	While our work awaits experimental confirmation in state-of-the-art experiments with 1D trapped quantum gases in the weak degeneracy regime, we plan to extend our analysis to momentum correlations in the 1D Bose gas that have been measured recently \cite{Fang2016}.
\begin{acknowledgements}
	B.G.\@ thanks the \textit{Studienstiftung des deutschen Volkes} for support. We	further	acknowledge	financial support through the Deutsche Forschungsgemeinschaft (through SFB 1277, Project A07). We thank Piotr Deuar and Peter Drummond for providing the numerical data from Ref.~\cite{Deuar2009}.
\end{acknowledgements}
\onecolumngrid
\vspace{0.3cm}
\begin{center}
	\rule{0.5\linewidth}{1px}
\end{center}
\vspace{0.3cm}
\newpage
\twocolumngrid
\appendix
\section{Pair correlation function in the canonical ensemble}
\label{app:canonical}
To derive Eq.~(\ref{eqn:g2}) from (\ref{eqn:g2general}) we perform the trace in $\braket{\mathcal{O}}=\operatorname{Tr}_+^{(N)}\{e^{-\beta \hamilton}\mathcal{O}\}/Z_+^{(N)}$ in the position basis of the Hilbert space of symmetric $N$-particle states
\begin{equation}
	|\vectx)=\frac{1}{N!}\sum_{P\in S_N}\ket{P\vectx}=\frac{1}{\sqrt{N!}}\hat{\Psi}^\dagger(\vectx_1)\dots\hat{\Psi}^\dagger(\vectx_N)|0)
\end{equation}
and insert a closure relation to rewrite Eq.~(\ref{eqn:g2}) as
\begin{multline}
	g_2^{(N)}(\vect{r})=\frac{1}{\rho^2 Z_+^{(N)}}\int \Diff{N}{\vectx}\Diff{N}{\vectx'}(\vectx'|e^{-\beta\hamilton}|\vectx)\\
	\times (\vectx|\hat{\Psi}^\dagger(0)\hat{\Psi}^\dagger(\vect{r})\hat{\Psi}(\vect{r})\hat{\Psi}(0)|\vectx').
\end{multline}
The first term in the integral is exactly the symmetry-projected many-body propagator $K_+^{(N)}(\vectx',\vectx;t=-\i\hbar\beta)$ defined in Eq.~(\ref{eqn:symmetric_propagator}), as the symmetry projection commutes with $\hamilton$ and is idempotent. The function
\begin{multline}
	C(\vectx,\vectx',\vect{r})
	=
	(\vectx|\hat{\Psi}^\dagger(0)\hat{\Psi}^\dagger(\vect{r})\hat{\Psi}(\vect{r})\hat{\Psi}(0)|\vectx')
	\\=
	\frac{1}{N!}(0|\hat{\Psi}(\vectx_N)\dots\hat{\Psi}(\vectx_1)\hat{\Psi}^\dagger(0)\hat{\Psi}^\dagger(\vect{r})\hat{\Psi}(\vect{r})\hat{\Psi}(0)
	\\
	\times	
	\hat{\Psi}^\dagger(\vectx_1')\dots\hat{\Psi}^\dagger(\vectx_N')|0)
\end{multline}
can be easily evaluated using Wick's theorem with the definition of a contraction $\hat{A}^\bullet\hat{B}^\bullet=\hat{A}\hat{B}-{}\NormalO{\hat{A}\hat{B}}$, where $\NormalO{\hat{A}\hat{B}}$ stands for normal ordering of the field operators $\hat{A},\hat{B}$. The only nonvanishing contractions are then
\begin{equation}
	\hat{\Psi}(\vect{y})^\bullet\hat{\Psi}^{\dagger}(\vect{z})^\bullet=\delta(\vect{y}-\vect{z}),
\end{equation}
and Wick's theorem states that $C(\vectx,\vectx',\vect{r})$ is given by the sum of all full contractions of all the field operators. For a nonvanishing contribution, $\hat{\Psi}(\vect{r})$ and $\hat{\Psi}(0)$ are contracted to the right while $\hat{\Psi}^\dagger(\vect{r})$ and $\hat{\Psi}^\dagger(0)$ are contracted to the left. As we can relabel the coordinates under the integral [$K_+^{(N)}(\vectx',\vectx;t=-\i\hbar\beta)$ is symmetric in $\vectx'$ and $\vectx$], all of the $N(N-1)N!$ nonvanishing contractions give exactly the same contribution to $g_2^{(N)}$ and we can replace
\begin{multline}
	C(\vectx,\vectx',\vect{r})\rightarrow N(N-1)\prod_{i=3}^{N}\delta(\vectx_i'-\vectx_i) \\
	\times \delta(\vectx_1)\delta(\vectx_1')\delta(\vectx_2-\vect{r})\delta(\vectx_2'-\vect{r})
\end{multline}
in the integral, which immediately gives Eq.~(\ref{eqn:g2}). Equation (\ref{eqn:Z}) is easily obtained from $\operatorname{Tr}_+^{(N)}\{e^{-\beta\hamilton}\}$.
\section{Derivation of the propagator for the extended Lieb-Liniger gas}
\label{app:propagator_derivation}
The symmetric wave functions of the continuum limit of the LL model are known and can be written as \cite{gaudin2014,korepin1997}
\begin{equation}
	\chi_\vect{k}(\vectx)=\frac{1}{\sqrt{(2\pi)^N N!}}\sum_{P\in S_N}(-1)^Pf(\hat{P}\vect{k},\vectx)\eh{\i(\hat{P}\vectk)\vectx}.
\end{equation}
Here, $S_N$ is the symmetric group acting on the index set, $\hat{P}$ is the $N\times N$ matrix representation of the permutation $P$ such that $(\hat{P}\vectk)_i=k_{P(i)}$, $(-1)^P$ is the sign of the permutation $P$, and
\begin{equation}
	f(\vect{k},\vectx)=\prod_{j>l}\frac{k_j-k_l-ic \sgn(x_j-x_l)}{\left[(k_j-k_l)^2+c^2\right]^{\frac{1}{2}}}.
	\label{eqn:app_f_function}
\end{equation}
They obey the  Schr\"{o}dinger equation
\begin{equation}
	\hamilton\chi_\vectk(\vectx)=\frac{\hbar^2\vectk^2}{2m}\chi_\vectk(\vectx)
\end{equation}
for the LL Hamiltonian defined in Eq.~(\ref{eqn:LL-hamiltonian}), but now with $x_i\in\R$. It has been proven in \cite{gaudin2014} that the wave functions $\chi_\vectk$ form a complete set in the domain $x_1<\dots<x_1$ if we choose $k_1<\dots<k_N$ and that they are normalized such that
\begin{equation}
	\int_{\R^N}\diff{\vectx}\chi_{\vectk'}(\vectx)\overline{\chi_\vectk(\vectx)}=\prod_j\delta(k_j'-k_j),
\end{equation}
where the bar denotes complex conjugation. The symmetric many-body propagator is thus given as
\begin{align}
	\K{N}_+(\vectx',\vectx;t)&=\int_{D}\diff{\vect{k}}\eh{-\frac{\i\hbar t}{2m}\vectk^2}\chi_\vectk(\vectx')\overline{\chi_\vectk(\vectx)}\\
	&=\frac{1}{N!}\int_{\R^N}\diff{\vectk}\eh{-\frac{\i\hbar t}{2m}\vectk^2}\chi_\vectk(\vectx')\overline{\chi_\vectk(\vectx)},
	\label{eqn:app_prop}
\end{align}
where $D$ is the domain with $k_1<\dots<k_N$. The second line follows from the fact that $\chi_\vectk(\vectx)$ is an antisymmetric function with respect to exchange of any two of the $k_i$ so that the integrand is a symmetric function ($\vectk^2$ is invariant under permutations). We would like to find a simplified form for this integrand. Using the transitivity of the symmetric group it is straightforward to show that
\begin{multline}
	\sum_{R,Q\in S_N}(-1)^{R\circ Q}f(\hat{R}\vectk,\vectx')\overline{f(\hat{Q}\vectk,\vectx)}\eh{\i(\hat{R}\vectk)\vectx'-\i(\hat{Q}\vectk)\vectx}\\
	=\sum_{P,Q\in S_N}(-1)^Pf(\hat{P}^{-1}\hat{Q}\vectk,\vectx')\overline{f(\hat{Q}\vectk,\vectx)}\eh{\i(\hat{Q}\vectk)(\hat{P}\vectx'-\vectx)},
	\label{eqn:app_intermediate}
\end{multline}
where we have substituted $R=Q\circ P^{-1}$. Note that the matrix representation of two successive permutations $R\circ S$ is $\hat{S}\hat{R}$, \ie, the order is reversed. The integrand in Eq.~(\ref{eqn:app_prop}) thus depends only on $\hat{Q}\vectk$ so that the sum over the permutations $Q$ gives just a factor of $N!$ as we can relabel the $k_i$ in each integral. The key step now is to realize that the function $f$ satisfies
\begin{equation}
	f(\hat{P}\vectk,\hat{P}\vect{x})=(-1)^Pf(\vectk,\vectx)
	\label{eqn:app_f_identity}
\end{equation}
for all permutations $P$. This will be proven at the end of the paragraph. A simple calculation then shows that
\begin{multline}
	(-1)^Pf(\hat{P}^{-1}\vectk,\vectx')\overline{f(\vectk,\vectx)}=f(\vectk,\hat{P}\vectx')\overline{f(\vectk,\vectx)}\\
	=\prod_{j>l}\frac{k_j-k_l-\i c\sgn(x_{P(j)}'-x_{P(l)}')}{k_j-k_l-\i c\sgn(x_j-x_l)}
\end{multline}
Putting everything together we find that the symmetric many-body propagator can be written as in Eq.~(\ref{eqn:LL-propagator}) with the effective many-body propagator (\ref{eqn:LL-eff_propagator}).

To complete the proof we still have to show the identity (\ref{eqn:app_f_identity}). The proof is trivial if we can show this for a permutation that interchanges only two successive numbers $m,m+1$ for $m=1,\dots,N-1$, as any permutation can be written as a composition of such exchange operations. The product of the denominators in the definition of $f$, Eq.~(\ref{eqn:app_f_function}), is invariant under permutations of the $k_i$. So we only have to consider the product of the numerators, that we can split [after fixing $m$ and $P=(m\quad m+1)$] into the factor where $j=m+1, l=m$
\begin{multline}
	k_{P(m+1)}-k_{P(m)}-\i c\sgn(x_{P(m+1)}-x_{P(m)})\\
	=-[k_{m+1}-k_m-\i c\sgn(x_{m+1}-x_m)]
	\label{eqn:app_first_factor}
\end{multline}
and all the other factors. We now have to prove that the product of the latter is invariant under $P$, as $(-1)^P=-1$ is already accounted for in the first factor (\ref{eqn:app_first_factor}). Let us define the set
\begin{equation}
	\Omega_P=\{k_{P(j)}-k_{P(l)}-\i c\sgn(x_{P(j)}-x_{P(l)})\vert j>l\},
\end{equation}
where we exclude the factor that has $j=m+1,l=m$. The proof is done if we show that $\Omega_P=\Omega_{id}$. Let us choose an element in $\Omega_P$. As $P$ interchanges the sign of $j-l$ if and only if both $j=m+1$ and $l=m$, it is also an element of $\Omega_{id}$. Together with the fact that both sets are of the same size, this shows their identity.

\section{Second- and third-order interaction contributions}
\label{app:dK3}
The second-order interaction contribution can easily be calculated from the propagator for a 1D $\delta$ potential $V(x)=(\hbar^2c/m)\,\delta(x)$ \cite{bauch1985,manoukian1989}:
\begin{equation}
K_\delta(x',x;t)=K_0(x',x;t)+K_c(x',x;t)
\end{equation}
with
\begin{equation}
	K_0(x',x;t)=\sqrt{\frac{m}{2\pi\i\hbar t}}\eh{-\frac{m}{2\i\hbar t}(x'-x)^2}
\end{equation}
and
\begin{equation}
K_c(x',x;t)=-\int_0^\infty\diff{u}\eh{-u}K_0\Big(|x'|+|x|+\frac{u}{c},0;t\Big).
\end{equation}
Introducing center of mass and relative coordinates this results in
\begin{equation}
\dK{2}(\vect{x}',\vect{x};t)=K_{0,M}(R',R;t)K_{c,\mu}(r',r;t),
\end{equation}
where the additional indices $M$ and $\mu$ stand for the total and reduced mass that should be used in the expressions.

The result for the third-order interaction contribution was calculated for the fundamental domain $\mathcal{F}$ defined as the region where $x_1<x_2<x_3$. The result for $\vectx$ or $\vectx'$ in another domain is then obtained by projecting both coordinates into $\mathcal{F}$ (\ie, ordering them by size). Expressing relative and center-of-mass coordinates in units of the thermal wavelength $\lambda_T=\sqrt{2\pi\hbar^2\beta/m}$ through
\begin{multline}
(\rt_1,\rt_2,\tilde{R})=\frac{\sqrt{2\pi}}{\lambda_T}(r_1,r_2,R)\\
=\frac{\sqrt{2\pi}}{\lambda_T}\left(\bar{x}_2-\bar{x}_1,\bar{x}_3-\bar{x}_2,\frac{\bar{x}_1+\bar{x}_2+\bar{x}_3}{3}\right),
\label{eqn:app_coordinates}
\end{multline}
where the bar denotes the projection to $\mathcal{F}$,
the simplified result in dimensionless coordinates and interaction parameter is\vspace{-.1cm}
\begin{widetext}
	\begin{align}
	\dK{3}_+&(\vectx',\vectx;t=-\i\hbar\beta)=\frac{1}{3\lambda_T^3}\exp\left[-\frac{3}{2}\left(\tilde{R}'-\tilde{R}\right)^2\right]
	\int_{0}^{\infty}\diff{u}\int_{-u}^{u}\diff{v}\nonumber\\
	\times
	\Bigg\{&
	\exp\left[
	-u-\frac{1}{4}\left(\rt_1-\rt_2'+\frac{v}{2\ct}\right)^2-\frac{1}{12}\left(\rt_1+\rt_2'+2(\rt_1'+\rt_2)+\frac{3u}{2\ct}\right)^2
	\right]
	+[\rt_i\leftrightarrow \rt_i']\nonumber\\
	-&
	\exp\left[
	-u-\frac{1}{4}\left(\rt_1+\rt_2'+\frac{v}{2\ct}\right)^2-\frac{1}{12}\left(\rt_1+\rt_2'+2(\rt_1'+\rt_2)+\frac{3u}{2\ct}\right)^2
	\right]
	+[\rt_i\leftrightarrow \rt_i']\nonumber\\
	+&
	3\exp\left[
	-u-\frac{1}{4}\left(\rt_1+\rt_2'+\rt_1'+\rt_2+\frac{u}{2\ct}\right)^2-\frac{1}{12}\left((\rt_1+\rt_2')-(\rt_1'+\rt_2)+\frac{3v}{2\ct}\right)^2
	\right]
	\Bigg\}.
	\label{eqn:dK3}
	\end{align}
\end{widetext}
Here, the interaction strength has been rescaled to $\ct=\lambda_T c/\sqrt{8\pi}$. The above result gets simplified if we are interested in the diagonal elements $\vectx'=\vectx$, as is the case for $b_3^{(2)}(r)$, where we have to set $x_1=0$, $x_2=r$ and integrate $x_3$ in Eq.~(\ref{eqn:dK3}) over full space. Due to the symmetry of the problem we can restrict ourselves to $r>0$ and thus have to consider three different regimes $x_3<0$, $0<x_3<r$, and $x_3>r$. This leads to different assignments of the variables $\tilde{r}_i$ in Eq.~(\ref{eqn:app_coordinates}) due to the projection onto the fundamental domain. By performing all the integrations and combining the result with the contributions from the diagrams of lower orders in the interaction contributions we get $b_3^{(2)}(r)=d_1(r)+d_2(r)+d_3(r)$ with
\begin{widetext}
	\begin{align}
	d_1\left(\frac{\lambda_T}{\sqrt{2\pi}}\rt\right)&=\sqrt{2}\eh{-\frac{3}{4}\rt^2},\label{eqn:d1}\\
	d_2\left(\frac{\lambda_T}{\sqrt{2\pi}}\rt\right)&=
	-2\sqrt{2}\eh{-\rt^2}\Big\{
	\eh{(\frac{\rt}{2})^2}\erfc\left(\frac{\rt}{2}\right)+(\rt\ct-1)\eh{(\ct+\frac{\rt}{2})^2}\erfc\left(\ct+\frac{\rt}{2}\right)
	\Big\}
	-8\sqrt{2}[F_{\ct}(\rt/2,-\rt/2)+F_{\ct}(0,\rt)],\label{eqn:d2}\\
	d_3\left(\frac{\lambda_T}{\sqrt{2\pi}}\rt\right)&=
	8\sqrt{2}\left\{
	F_{\ct}(0,\rt)
	+\left[2+2\ct\rt+\frac{8}{3}\ct^2\right]F_{\ct}(\rt/2,\rt/2)
	-\left[2\rt\ct+\frac{8}{3}\ct^2\right]F_{\ct}(\rt/2,\rt/2)
	+[2\rt\ct-1]G_{\ct}(\rt)\right. \nonumber\\
	&\quad
	\left. +\frac{4}{3}\ct^2\eh{-\rt^2}
	\left[
	\eh{(\frac{\rt}{2})^2}\erf\left(\frac{\rt}{2}\right)+\eh{(\ct+\rt)^2}\erfc(\ct+\rt)-\eh{(\ct+\frac{\rt}{2})^2}\erfc\left(\ct+\frac{\rt}{2}\right)
	\right]
	\right\}.\label{eqn:d3}
	\end{align}
\end{widetext}
The functions $F_{\ct}$ and $G_{\ct}$ are defined as
\begin{align}
& F_{\ct}(x,y)=\ct\int_{0}^{\infty}\diff{u}\eh{-4\ct u-3(u+x)^2}\erfc(u+y),\\
& G_{\ct}(x)=\sqrt{3}\ct\int_{0}^{\infty}\diff{u}\eh{-4\ct u-(u+x)^2}\erf(\sqrt{3}u).
\end{align}
The indices $n$ of the functions $d_n$ stand for the order of the interaction contributions that are involved, such that, \eg, $d_1$ is the result for free bosons. The function $b_3=b_3^{(0)}$ is obtained from $b_3^{(2)}(r)$ by using Eq.~(\ref{eqn:Bn_reduction}). As many of the resulting terms from second- and third-order interaction contributions cancel after integration, we present here only the sum of all contributions given by
\begin{equation}
b_3=\frac{1}{\sqrt{3}}+\frac{3}{2}\sqrt{3}\left[\eh{(2\ct)^2}\erfc(2\ct)-\tilde{F}_{\frac{1}{\sqrt{3}}}(\ct)-\tilde{F}_{\sqrt{3}}(\ct)\right]
\label{eqn:b3}
\end{equation}
with
\begin{equation}
\tilde{F}_\nu(\ct)=\frac{2}{\sqrt{\pi}}\eh{(1+\nu^2)\ct^2}\int_0^\infty\diff{u}\eh{-(u+\sqrt{1+\nu^2}\ct)^2}\erfc(\nu u).
\end{equation}
Note that $\tilde{F}_\nu(0)=1-\pi/2\arctan(\nu)$, which can easily be proven by differentiating $\tilde{F}_\nu(0)$ with respect to $\nu$, so that $\tilde{F}_{\nu}(0)+\tilde{F}_{\nu^{-1}}(0)=1$ for $\nu>0$ and thus $b_3=1/\sqrt{3}$ for $\ct=0$.

\section{Numerical calculation and error estimates}
\label{app:numerics}
\subsection{Numerical scheme for calculation of the pair-correlation function}
For the numerical calculation of the nonlocal pair correlation function we used the Bethe ansatz solutions \cite{gaudin2014}
\begin{equation}
	\chi_\vect{k}(\vectx)=C(\vectk)\sum_{P\in S_N}(-1)^Pf(\hat{P}\vect{k},\vectx)\eh{\i(\hat{P}\vectk)\vectx},
\end{equation}
with the function $f$ defined in (\ref{eqn:app_f_function}) and a (real) normalization constant $C(\vectk)$ that depends on the quasimomenta that solve the coupled transcendental equations
\begin{equation}
	\eh{\i k_j L}=-\prod_{i=1}^N\frac{k_j-k_i+\i c}{k_j-k_i-\i c},\qquad j=1,\dots,N.
	\label{eqn:app_transcendental}
\end{equation}
We focus only on the case $c>0$. In this case the logarithm of the equations (\ref{eqn:app_transcendental}) can be taken directly and the quasimomenta are determined by a set of $N$ ordered quantum numbers that represent the branch of the logarithm that is used in the respective equation (for more details see, \eg, \cite{lieb1963,korepin1997}). The solution is then easily found via Newton's method. The energy of the eigenstate $\chi_\vectk$ is given by
\begin{equation}
	E(\vectk)=\frac{\hbar^2\vectk^2}{2m}.
\end{equation}
The nonlocal pair-correlation function can now be written as
\begin{equation}
	g_2^{(N)}(r)=\frac{N-1}{N}\frac{1}{Z^{(N)}}\sum_{k_1<\dots<k_N}\eh{-\beta E(\vectk)}g_{2,\vectk}^{(N)}(r),
	\label{eqn:app_g2_exact}
\end{equation}
with
\begin{equation}
	g_{2,\vectk}^{(N)}(r)=L^2\int_0^L\diff{x_3}\dots\diff{x_N}|\chi_\vectk(0,r,x_3,\dots,x_N)|^2.
\end{equation}
The absolute square of the wave functions involves $(N!)^2$ terms and we could now integrate them directly as was done in \cite{Zill2016}. But we can reduce the problem to $N!$ such integrations with the help of similar manipulations as we have used in Appendix~\ref{app:propagator_derivation}. This enables us to write the absolute square of the wave function as
\begin{equation}
	|\chi_\vectk(\vectx)|^2=C^2(\vectk)\sum_{Q\in S_N}\Psi_{\hat{Q}\vectk}(\vectx),
\end{equation}
where $\hat{Q}$ is the matrix representation of the permutation $Q$ and
\begin{multline}
	\Psi_\vectk(\vectx)=\sum_{P\in S_N}\Re\Big\{(-1)^P\prod_{j>l}\frac{k_{P(j)}-k_{P(l)}-\i c}{k_j-k_l-\i c}\\
	\times \eh{\i(\hat{P}\vectk-\vectk)\mathcal{P_F}(\vectx)}\Big\}.
\end{multline}
Here, $\mathcal{P_F}(\vectx)$ is the projection of $\vectx$ to the fundamental domain $\mathcal{F}$ with $x_1<\dots<x_N$ and we can take the real part as the imaginary parts have to vanish in the overall sums. The plane waves have to be integrated over full space in $x_3,\dots,x_N$, which leads to different projections in the fundamental domain. Due to the translational invariance and inversion symmetry we can restrict ourselves to the cases where $0<\dots<x_{j}<r<x_{j+1}<\dots <L$. We implemented a simple algorithm that correctly traces out $x_3$ to $x_N$ symbolically. The final expression that we used for the numerical calculation is
\begin{equation}
	g_{2,\vectk}^{(N)}(r)=L\frac{\sum_{Q\in S_N}F_{\hat{Q}\vectk}(r)}{\sum_{Q\in S_N}G_{\hat{Q}\vectk}},
\end{equation}
with the functions
\begin{align}
	F_\vectk(r)&=\sum_{P\in S_N}\Re\left\{f_\vectk^{(P)}h_\vectk^{(P)}(r)\right\},\\
	G_\vectk&=\sum_{P\in S_N}\Re\left\{f_\vectk^{(P)}\int_{0}^{L}\diff{r}h_\vectk^{(P)}(r)\right\},
\end{align}
where we defined
\begin{align}
	f_\vectk^{(P)}&=(-1)^P\prod_{j>l}\frac{k_{P(j)}-k_{P(l)}-\i c}{k_j-k_l-\i c},\\
	h_\vectk^{(P)}(r)&=\int_{0}^{L}\diff{x_3}\dots\diff{x_N}\eh{\i(\hat{P}\vectk-\vectk)\mathcal{P_F}(\vectx)}\big|_{x_1=0,x_2=r}.
\end{align}
One has to take care, as some of the sets of $k_i$ obey certain symmetries leading to divergencies in the symbolic expressions for $h_{\hat{Q}\mathbf{k}}^{(P)}$ for certain permutations $Q$.
These cases have to be treated separately, leading to a piecewise definition of the functions $G_\vectk$ and $F_\vectk$ with respect to $\vectk$. For computation the length $L$ of the system can be completely eliminated by rescaling the variables according to $\vectk\mapsto \vectk L,E\mapsto EL^2, x\mapsto x/L, c\mapsto cL,\beta\mapsto\beta/L^2$.
One may note that for $N=3$ particles the integral of a plane wave can be written
\begin{equation}
	\int_{a}^{b}\diff{x}\eh{\i\kappa x}=(b-a)\eh{\i\frac{b+a}{2}\kappa}\operatorname{sinc}\left(\frac{b-a}{2}\kappa\right),
\end{equation}
which is well defined for all values of $\kappa$ and thus we can use this integral for $\kappa=k_{P(i)}-k_i$ for all permutations $P$ and the respective indices $i$.

\subsection{Error estimation}
We need to find an estimate for the error that occurs if we truncate the summation over quasi-momenta to a certain cutoff energy. We therefore write the nonlocal pair correlation function as
\begin{equation}
	g_2^{(N)}(r)=\frac{N-1}{N}\frac{A^{(N)}(r)}{Z^{(N)}},
\end{equation}
with $A^{(N)}$ defined by Eq.~(\ref{eqn:app_g2_exact}), and we denote a cutoff in the energy by a bar at the respective quantities. Both $\bar{A}^{(N)}(r)$ and $\bar{Z}^{(N)}$ are positive and monotonously increasing with the cutoff energy. Let us write $\bar{A}^{(N)}(r)=[1-\epsilon_A(r)]A^{(N)}(r), \bar{Z}^{(N)}=(1-\epsilon_Z)Z^{(N)}$ with the positive relative errors $\epsilon_A(r)$ and $\epsilon_Z$. The relative error of $\bar{g}_2^{(N)}(r)$ is then
\begin{equation}
	\epsilon_g(r)=\frac{1-\epsilon_A(r)}{1-\epsilon_Z}-1=[\epsilon_Z-\epsilon_A(r)][1+\mathcal{O}(\epsilon_Z)].
\end{equation}
Using the normalization
\begin{equation}
	\int_{0}^{L}\diff{r}A^{(N)}(r)=LZ^{(N)},
\end{equation}
which also holds for the truncated objects, it is easily shown that the absolute error of $\bar{g}_2^{(N)}(r)$ averages out,
\begin{equation}
	\int_{0}^{L}\diff{r}\epsilon_g(r)g_2^{(N)}(r)=0.
\end{equation}
We can now define
\begin{align}
	\epsilon_g^>(r)&=
	\begin{cases}
		\epsilon_g(r)& \epsilon_g(r)>0\\
		\hspace{.8em}0& \text{else}
	\end{cases}\\
	\epsilon_g^<(r)&=\epsilon_g(r)-\epsilon_g^>(r).
\end{align}
As $\epsilon_A(r)$ is positive $\epsilon_g^>(r)$ is bound from above by $\epsilon_Z[1+\mathcal{O}(\epsilon_Z)]$ and we have
\begin{multline}
	\frac{1}{L}\int_{0}^{L}\diff{r}\epsilon_g^>(r)g_2^{(N)}(r)=\frac{1}{L}\int_{0}^{L}\diff{r}|\epsilon_g^<(r)|g_2^{(N)}(r)\\
	<\frac{N-1}{N}\epsilon_Z[1+\mathcal{O}(\epsilon_Z)]
\end{multline}
for the absolute error of $\epsilon_g(r)g_2^{(N)}$. Thus, even though the latter could, in principle, take large negative values down to $\epsilon_g(r) g_2^{(N)}(r)=-g_2^{(N)}(r)$ at certain points, this can only be the case in a small region that scales with the inverse of this value and with $\epsilon_Z$, meaning that the absolute error is smaller than $\epsilon_Z$ everywhere else. However, we do not expect (and do not observe) such peaked drops in the pair correlation-function, as they can be regarded as unphysical.

In order to have an estimate for the relative error $\epsilon_Z$ of $Z^{(N)}$ we use the observation that the mean density of states $\rho^{(N)}(E,c)$ in the LL model obeys $\rho^{(N)}(E,0)\geq\rho^{(N)}(E,c)\geq\rho^{(N)}(E,\infty)$ and use the two limits for an estimate of the error in $Z^{(N)}$. As we are mainly interested in the approximation error for high temperatures, where the sum over the exact states in the partition function converges slowly, we can make use of the semiclassical approximations. The mean density of states is given by the inverse Laplace transform with respect to $\beta$ of the semiclassical partition function $Z^{(N)}$, Eq.~(\ref{eqn:Zrecursion}). For the limits of free bosons and fermionization this can be written as \cite{hummel2013}
\begin{align}
	Z_\pm^{(N)}(\beta)&=\frac{1}{N!}\sum_{l=1}^{N}(\pm1)^{N-l}z_l^{(N)}\left(\frac{L}{\lambda_T}\right)^l\nonumber\\
	&=\frac{1}{N!}\sum_{l=1}^{N}(\pm1)^{N-l}z_l^{(N)}\left(\frac{\alpha}{\beta}\right)^{\frac{l}{2}},
\end{align}
with $\alpha=mL^2/(2\pi\hbar^2)$, where the sign stands for the limits of free bosons $(+)$ and fermionization $(-)$, respectively. The numbers $z_l^{(N)}$ contain the sum of diagrams corresponding to the partitions of $N$ particles into $l$ clusters and are independent of the temperature, as $b_n=(\pm 1)^{n-1}/\sqrt{n}$ for the two limits. The mean density of states is
\begin{align}
	\rho_\pm^{(N)}(E)
	&=\invLaplace{\beta}{Z_\pm^{(N)}(\beta)}{E}\nonumber\\
	&=\frac{1}{N!}\sum_{l=1}^{N}(\pm1)^{N-l}z_l^{(N)}\alpha^{\frac{l}{2}}\frac{E^{\frac{l}{2}-1}}{\Gamma(\frac{l}{2})},
\end{align}
with the gamma function $\Gamma(x)$. The relative error in the partition function is then approximated by the semiclassical error
\begin{align}
	\tilde{\epsilon}_{Z_\pm}(x,\beta)
	&=\frac{1}{Z_\pm^{(N)}}\int_{x/\beta}^{\infty}\diff{E} \rho_\pm^{(N)}(E)\eh{-\beta E}\nonumber\\
	&=\frac{
	\sum_{l=1}^{N}z_l^{(N)}\left(\pm\frac{\lambda_T}{L}\right)^{N-l}Q(\frac{l}{2},x)
	}{
	\sum_{l=1}^{N}z_l^{(N)}\left(\pm\frac{\lambda_T}{L}\right)^{N-l},
	}
	\label{eqn:app_semiclassical_error}
\end{align}
where $Q(a,x)$ is the regularized incomplete gamma function
\begin{equation}
	Q(a,x)=\frac{\Gamma(a,x)}{\Gamma(a)}=\frac{\int_{x}^{\infty}\diff{t}t^{a-1}\eh{-t}}{\int_{0}^{\infty}\diff{t}t^{a-1}\eh{-t}}.
\end{equation}
\vspace{0em}

We are interested in the regime $\lambda_T\leq 0.5 L$ and, for reasonably small errors, $x\gtrsim 10$. The Numerator in Eq.~(\ref{eqn:app_semiclassical_error}) is then dominated by the $l=N$ term, and the error is largest if we minimize the denominator by using the result for the fermionization limit. This may also be seen from the fact that the ground-state energy is maximized in this limit, maximizing the ratios $\eh{-\beta(E_k-E_0)}$ in $Z=\eh{-\beta E_0}(1+\eh{-\beta(E_1-E_0)}+\dots)$. We thus used the semiclassical error estimate in the fermionization limit as a bound for the error at arbitrary couplings. In our numerical calculations we have used the cutoff $x=20$ for $N=3,4$, leading to $\tilde{\epsilon}_{Z_-}<4\times 10^{-8}$ and $\tilde{\epsilon}_{Z_-}<8\times 10^{-7}$, respectively (for all temperatures). For $N=5$ we have used $x=14$ for $\lambda_T=0.1L$ ($\tilde{\epsilon}_{Z_-}<6.2\times10^{-5}$, approximately  $1.4\times 10^5$ to $2.5\times 10^5$ states) and $x=22$ for $\lambda_T=0.4L$ ($\tilde{\epsilon}_{Z_-}<8.2\times 10^{-7}$, approximately 250 to 1330 states), respectively, in the corresponding semiclassical approximation.

\bibliography{bibliography}
\end{document}